\documentclass{book}    
\usepackage{piers}  
\begin{document}

\title{Some Wave Equations for Electromagnetism and Gravitation}
\maketitle

\author      {Zi-Hua Weng}
\affiliation {Xiamen University}
\address     {}
\city        {Xiamen}
\postalcode  {}
\country     {China}
\phone       {345566}    
\fax         {233445}    
\email       {xmuwzh@xmu.edu.cn}  
\misc        { }  
\nomakeauthor

\begin{authors}

{\bf Zi-Hua Weng}\\
\medskip
School of Physics and Mechanical \& Electrical Engineering,
\\Xiamen University, Xiamen 361005, China\\

\end{authors}

\begin{paper}

\begin{piersabstract}
The paper studies the inferences of wave equations for electromagnetic fields when there are gravitational fields at the same time. In the description with the algebra of octonions, the inferences of wave equations are identical with that in conventional electromagnetic theory with vector terminology. By means of the octonion exponential function, we can draw out that the electromagnetic waves are transverse waves in a vacuum, and rephrase the law of reflection, Snell's law, Fresnel formula, and total internal reflection etc. The study claims that the theoretical results of wave equations for electromagnetic strength keep unchanged in the case for coexistence of gravitational and electromagnetic fields. Meanwhile the electric and magnetic components of electromagnetic waves can not be determined simultaneously in electromagnetic fields.
\end{piersabstract}


\psection{Introduction}

In the electromagnetic field, the inferences of wave equation are believed to be correct. Until to now, this validity is confined only to one special condition that there exist only the electromagnetic fields. When there are gravitational fields, the people doubt whether the wave equation of electromagnetic field is still correct or not. The validity about wave equations is being questioned continuously, and it remains as puzzling as ever. The existing theories do not explain why the deductions of wave equation should keep unchanged, and then do not offer compelling reason for the unique situation. The paper attempts to find out why the deductions of wave equation keep the same in most cases, even in the gravitational field.

The law of reflection (or specular reflection) is the mirror-like reflection of light from a surface, and the angle of incidence equals the angle of reflection \cite{paasschens, polkovnikov}. This behavior was first discovered through careful observation and measurement by Heron of Alexandria, Greek. The law of reflection was dealt with the Principle of the Shortest Path of Light, which was later stated in this form by P. Fermat in 1662 \cite{sundar}. The most modern form is that the path is at an extremum \cite{merano, redkov}.

Snell's law (also known as law of refraction) was first accurately described in a mathematical form by Ibn Sahl of Baghdad. It was rediscovered by T. Harriot in 1602, who did not publish his results. In 1621, W. Snell derived a mathematically equivalent form, that remained unpublished during his lifetime, until C. Huygens mentioned Snell's discovery in 1678 \cite{buenker, ghosh}.

A.-J. Fresnel rediscovered interference and provided a mathematical theory of diffraction based on the wave theory \cite{qin, schomerus}. In 1817, T. Young proposed a small transverse component to light, while yet retaining a far larger longitudinal component. By the year 1821, Fresnel was able to show via mathematical methods that polarization could be explained only if light was entirely transverse, with no longitudinal vibration whatsoever. He relied on mathematical analysis which enabled him to make predictions which could be accurately verified and he removed many of the objections to the wave theory of light \cite{skaar, obukhov}.

The algebra of quaternions was invented by W. R. Hamilton in 1843 \cite{hamilton}, and then was first used by J. C. Maxwell to represent field equations of electromagnetic field in 1861 \cite{maxwell}. O. Heaviside in 1884 recast Maxwell equation in terms of vector terminology and electromagnetic forces \cite{rothman}, thereby reduced the original twenty equations down to the four differential equations.

In 1871, H. Helmholtz made clear the electromagnetic theory effectively \cite{baruch, moroz}. And Helmholtz equation and boundary conditions can deduce some wave features, including the law of reflection, Snell's law, Fresnel formula, etc \cite{ben}. Later H. R. Hertz became famous as the first to demonstrate the electromagnetic radiation.

In the paper, in terms of the quaternion feature, we obtain field equation and related wave feature in the gravitational field. Rephrasing with the algebra of octonions, we achieve Maxwell equation in electromagnetic fields, and then some wave properties including the wave equation, law of reflection, Snell's law, Fresnel formula, and total internal reflection, etc.

\psection{Wave equation for field strength in gravitational field}

The wave feature of gravitational field can be described by the algebra of quaternions. In the quaternion space, the coordinates are $r_0$, $r_1$, $r_2$, and $r_3$, with the basis vector $\mathbb{E}_g = (1 , \emph{\textbf{i}}_1 , \emph{\textbf{i}}_2 , \emph{\textbf{i}}_3)$. The radius vector is $\mathbb{R}_g = \Sigma (r_i \emph{\textbf{i}}_i)$, and the velocity $\mathbb{V}_g = \Sigma (v_i \emph{\textbf{i}}_i)$, with $\emph{\textbf{i}}_0 = 1 $. Herein
$r_0 = v_0 t$ . $v_0$ is the speed of gravitational intermediate boson, and $t$ is the time. $i = 0, 1, 2, 3 $. $j = 1, 2, 3 $.

In the quaternion space for the gravitational field, the gravitational potential is, $\mathbb{A}_g = \Sigma (a_i \emph{\textbf{i}}_i)$. And the gravitational strength is $\mathbb{B}_g = \Sigma (b_i \emph{\textbf{i}}_i) = \lozenge \circ \mathbb{A}_g$ .The gravitational strength $\mathbb{B}_g$ covers two components, $\textbf{g}/v_0 = \partial_0 \textbf{a} + \nabla a_0 $ , and $ \textbf{b} = \nabla \times \textbf{a}$ . The gauge equation is $b_0 = \partial_0 a_0 + \nabla \cdot \textbf{a} = 0$, with $\textbf{a} = \Sigma (a_j \emph{\textbf{i}}_j)$ and $\nabla = \Sigma (\emph{\textbf{i}}_j \partial_j)$ . Herein the $\circ$ denotes the quaternion multiplication, the operator $ \lozenge = \Sigma (\emph{\textbf{i}}_i \partial_i)$, with $\partial_i = \partial/\partial r_i$ .

Comparing with classical gravitational theory, the linear momentum density, $\mathbb{P}= m \mathbb{V}_g$, is the source $\mathbb{S}_g$ of gravitational field. And the latter one is defined as follows,
\begin{eqnarray}
( \mathbb{B}_g / v_0 + \lozenge)^* \circ \mathbb{B}_g = - \mu_g \mathbb{S}_g~,
\end{eqnarray}
where $m$ is the mass density. $*$ denotes the quaternion conjugate. $\mu_g = 4 \pi G / v_0^2$ is a coefficient, and $G$ is the gravitational constant. $\mathbb{B}_g^* \circ \mathbb{B}_g / (2 \mu_g)$ is the gravitational field energy density, with the extra mass part $\triangle m = \mathbb{B}_g^* \circ \mathbb{B}_g / (v_0^2 \mu_g)$.

Further, the above can be separated into
\begin{eqnarray}
& \nabla \cdot \textbf{b} = 0~, ~\partial_0 \textbf{b} + \nabla^* \times \textbf{g} / v_0 = 0~,
\\
& \nabla^* \cdot \textbf{g} = - \widehat{m} / \varepsilon_g~, ~ \partial_0 \textbf{g} / v_0 + \nabla^* \times \textbf{b} = - \mu_g
\textbf{s}~,
\end{eqnarray}
where the coefficient $\varepsilon_g = 1/(\mu_g v_0^2)$. $\textbf{s} = \Sigma (s_j \emph{\textbf{i}}_j) $. $s_j = p_j = m v_j $. $s_0 = p_0 = \widehat{m} v_0$. $ \widehat{m} = m + \triangle m $.

The above means that the Newton's law of gravitation in the quaternion space \cite{weng} is the same as that in classical gravitational theory \cite{newton, wu}.

\begin{table}[h]
\caption{The quaternion multiplication table.} \label{tab:table1}
\centering
\begin{tabular}{ccccc}
\hline\hline $ $ & $1$ & $\emph{\textbf{i}}_1$  &
$\emph{\textbf{i}}_2$ & $\emph{\textbf{i}}_3$  \\
\hline $1$ & $1$ & $\emph{\textbf{i}}_1$  & $\emph{\textbf{i}}_2$ &
$\emph{\textbf{i}}_3$    \\
$\emph{\textbf{i}}_1$ & $\emph{\textbf{i}}_1$ & $-1$ &
$\emph{\textbf{i}}_3$  & $-\emph{\textbf{i}}_2$  \\
$\emph{\textbf{i}}_2$ & $\emph{\textbf{i}}_2$ &
$-\emph{\textbf{i}}_3$ & $-1$ & $\emph{\textbf{i}}_1$  \\
$\emph{\textbf{i}}_3$ & $\emph{\textbf{i}}_3$ &
$\emph{\textbf{i}}_2$ & $-\emph{\textbf{i}}_1$ & $-1$ \\
\hline\hline
\end{tabular}
\end{table}

\psubsection{Quaternion exponential function}

In the quaternion space, defining the complex quaternion as
\begin{eqnarray}
exp \left\{ \textbf{i} ( a + i b ) \right\} = cos ( a + i b ) + \textbf{i} sin ( a + i b )~,
\end{eqnarray}
where $a$ and $b$ are all real. $i$ is the imaginary unit, with $i^2 = -1 $ . Meanwhile the quaternion unit $\textbf{i}$ is one function of $\emph{\textbf{i}}_j$ , with $ \textbf{i} \circ \textbf{i} = -1$.

When $b = 0$ , the above reduces to the quaternion, $ exp ( \textbf{i} a ) = cos ( a ) + \textbf{i} sin ( a )$ . When $a = 0$ , Eq.(4) reduces to the Hamiltonian bi-quaternion,
\begin{eqnarray}
exp ( \textbf{i} i b ) = cos ( i b ) + \textbf{i} sin ( i b ) ~.
\end{eqnarray}

The bi-quaternion $exp(-i \textbf{i} \alpha)$ satisfies,
\begin{eqnarray}
& exp(-i \textbf{i} \alpha) \circ exp(i \textbf{i} \alpha) = 1~,~ exp(-i \textbf{i} \alpha) \circ exp(-i \textbf{i} \alpha) = exp(-i \textbf{i} 2\alpha)~,
\\
& \partial_j exp(-i \textbf{i} \alpha) = i \textbf{i} k_j \circ exp(-i \textbf{i} \alpha) ~, ~ \partial_j^2 exp(-i \textbf{i} \alpha) = k_j^2 exp(-i \textbf{i} \alpha) ~,
\end{eqnarray}
where $\alpha = - \Sigma (k_j r_j)$, with $k_j$ being the real coefficient.

In case of neglecting the direction of $\textbf{i}$ , Eq.(4) reduces to the complex exponential function,
\begin{eqnarray}
exp \left\{ i ( a + i b ) \right\} = cos ( a + i b ) + i sin ( a + i b ) = exp ( i  a ) exp ( - b ) ~ .
\end{eqnarray}

\begin{table}[h]
\caption{The operator and multiplication of the physical quantity in
the quaternion space.}
\label{tab:table2}
\centering
\begin{tabular}{ll}
\hline\hline
definition                  &   meaning                                              \\
\hline
$\nabla \cdot \textbf{a}$   &  $-(\partial_1 a_1 + \partial_2 a_2 + \partial_3 a_3)$ \\
$\nabla \times \textbf{a}$  &  $\emph{\textbf{i}}_1 ( \partial_2 a_3
                                 - \partial_3 a_2 ) + \emph{\textbf{i}}_2 ( \partial_3 a_1
                                 - \partial_1 a_3 ) + \emph{\textbf{i}}_3 ( \partial_1 a_2
                                 - \partial_2 a_1 )$                                 \\
$\nabla a_0$                &  $\emph{\textbf{i}}_1 \partial_1 a_0
                                 + \emph{\textbf{i}}_2 \partial_2 a_0
                                 + \emph{\textbf{i}}_3 \partial_3 a_0  $             \\
$\partial_0 \textbf{a}$     &  $\emph{\textbf{i}}_1 \partial_0 a_1
                                 + \emph{\textbf{i}}_2 \partial_0 a_2
                                 + \emph{\textbf{i}}_3 \partial_0 a_3  $             \\
\hline\hline
\end{tabular}
\end{table}

\psubsection{Wave equation}

Researching the transmission problem of gravitational waves needs to solve field equation. In a vacuum far away from the gravitational sources, there does not exist the mass $s_0$ and linear momentum $\textbf{s}$, and then field equation can be reduced to,
\begin{eqnarray}
\lozenge^* \circ \mathbb{B}_g = 0~.
\end{eqnarray}

Applying the operator $\lozenge$ to the above, we get the wave equation
\begin{eqnarray}
\lozenge \circ (\lozenge^* \circ \mathbb{B}_g) = 0~,~or~ (\partial_0^2 + \partial_1^2 + \partial_2^2 + \partial_3^2) \mathbb{B}_g = 0~.
\end{eqnarray}

The above is the Laplace equation in the Euclid space, meanwhile it is the wave equation in the quaternion space. And this Laplace equation can also be obtain from Eq.(9) by means of the conjugate operator $\lozenge^*$ and the field equation.

Proceeding with the operator $\partial_0$ and $\nabla$, we can obtain the wave equation about the components of field strength from the field equation Eqs.(2) and (3) directly,
\begin{eqnarray}
 (\partial_0^2 + \partial_1^2 + \partial_2^2 + \partial_3^2) \textbf{g} = 0~,
~ (\partial_0^2 + \partial_1^2 + \partial_2^2 + \partial_3^2) \textbf{b} = 0~.
\end{eqnarray}

The above means that Eq.(10) can be separated into two parts in Eq.(11) when $\mathbb{S}_g = 0$. And two strength components of gravitational field, $\textbf{b}$ and $\textbf{g}$, are both possessed of wave features, although their wave impact may be too tiny to be detected at present.

\psubsection{Transverse wave}

For the gravitational wave with the angular frequency $\omega$, the field strength should be a harmonic function, $cos \omega t$. In order to operate handily and cover more physics contents, it is convenient to substitute the function $exp(-i \omega t)$ for the $cos \omega t$.

The gravitational strength $\textbf{g}$ and $\textbf{b}$ can be written as follows,
\begin{eqnarray}
\textbf{g} = \textbf{g}(r) exp(- i \omega t ) ~,~ \textbf{b} = \textbf{b}(r) exp(- i \omega t ) ~,
\end{eqnarray}
combining this with Eq.(11), gives
\begin{eqnarray}
\left\{ - ( \omega / v_0 )^2 + \Sigma \partial_j^2 \right\} \textbf{g}(r) = 0~, ~ \left\{ - ( \omega / v_0 )^2 + \Sigma \partial_j^2 \right\} \textbf{b}(r) = 0~.
\end{eqnarray}

From the analysis of field equation, we find that the amplitude of gravitational strength will be increased steadily or deceased continuously. And then the field strength should be one hyperbolic cosine, $cos ( i \alpha)$. It will substitute the function $exp(-i \textbf{i} \alpha)$ for the $cos ( i \alpha)$. Herein the wave vector $\textbf{k} = \Sigma (\emph{\textbf{i}}_j k_j)$, while the vector radius $\textbf{r} = \Sigma (\emph{\textbf{i}}_j r_j)$, and $i \alpha$ is an imaginary angle.

And then, the gravitational strength $\textbf{g}(r)$ and $\textbf{b}(r)$ are
\begin{eqnarray}
\textbf{g}(r) = \textbf{g}_0 \circ exp(-i \textbf{i} \alpha) ~,
~ \textbf{b}(r) = \textbf{b}_0 \circ exp(-i \textbf{i} \alpha) ~,
\end{eqnarray}
where $\textbf{g}_0$ and $\textbf{b}_0$ both are constant vectors in the quaternion space.

Further, substituting the above in Eq.(13), we have the result,
\begin{eqnarray}
- (\omega / v_0 )^2 + \Sigma k_j^2 = 0~.
\end{eqnarray}

In the gravitational field, we will find that gravitational waves are the transverse waves in a vacuum. From the field equations
\begin{eqnarray}
\nabla \cdot \textbf{g} = 0~,~\nabla \cdot \textbf{b} = 0~,
\end{eqnarray}
we obtain
\begin{eqnarray}
\textbf{k} \cdot \textbf{g}_0 = 0~, ~\textbf{k} \cdot \textbf{g}'_0 = 0~, ~\textbf{k} \cdot \textbf{b}_0 = 0~, ~\textbf{k} \cdot \textbf{b}'_0 = 0~,
\end{eqnarray}
where $\textbf{g}'_0 = \textbf{g}_0 \circ \textbf{i}$ , $\textbf{b}'_0  =  \textbf{b}_0 \circ \textbf{i}$ .

The above states that the gravitational wave components, $\textbf{g}_0$ and $\textbf{b}_0$ , belong to the transverse waves. So do these two new wave components, $\textbf{g}'_0$ and $\textbf{b}'_0$ . Moreover, the amplitudes of wave components, $\textbf{g}'_0$ and $\textbf{b}'_0$ , are the same as that of $\textbf{g}_0$ and $\textbf{b}_0$ respectively.

In the gravitational field, there are two new kinds of gravitational waves. The field equations,
\begin{eqnarray}
\nabla \times \textbf{g} = \partial \textbf{b} / \partial t~,~\nabla \cdot \textbf{g} = 0~,
\end{eqnarray}
can be rewritten as,
\begin{eqnarray}
\nabla \circ \textbf{g} = \partial \textbf{b} / \partial t~.
\end{eqnarray}

Expanding the $exp(-i \textbf{i} \alpha)$, the above implies that
\begin{eqnarray}
[ \textbf{k} \circ \textbf{g}'_0 + \omega \textbf{b}_0 ] cosh (\alpha )
+ [ \textbf{k} \circ \textbf{g}_0 - \omega \textbf{b}'_0 ] sinh (\alpha ) = 0~,
\end{eqnarray}
considering Eq.(17) and the $\alpha$ can be an arbitrary value, and then
\begin{eqnarray}
\textbf{k} \times \textbf{g}'_0 + \omega \textbf{b}_0 = 0~,
~\textbf{k} \times \textbf{g}_0 - \omega \textbf{b}'_0 = 0~,
\end{eqnarray}
where $sin ( i \alpha)$ = $ i sinh (\alpha)$ , $cos ( i \alpha)$ = $cosh(\alpha)$ .

In the same way, from the field equations
\begin{eqnarray}
v_0^2 \nabla \times \textbf{b} = \partial \textbf{g} / \partial t~,~\nabla \cdot \textbf{b} = 0~,
\end{eqnarray}
we get
\begin{eqnarray}
\textbf{k} \times \textbf{b}'_0 + \omega (\textbf{g}_0 / v_0) / v_0 = 0~,
~\textbf{k} \times \textbf{b}_0 - \omega (\textbf{g}'_0 / v_0) / v_0 = 0~.
\end{eqnarray}

The above means that there are some relationships between $\mathbb{B}_g$ with $\mathbb{B}_g \circ \textbf{i}$ . The $\textbf{k}$ and $\textbf{g}_0$ will yield a new component $\textbf{b}'_0$ , while the $\textbf{k}$ and $\textbf{b}_0$ produce the $\textbf{g}'_0$ . The $\textbf{g}_0$ and $\textbf{b}'_0$ can be determined at the same time, while the $\textbf{b}_0$ and $\textbf{g}'_0$ can be measured simultaneously. Provided there are some substances in the nature may slow down the speed of gravitational waves, the field strength will possess the wave features, including the reflection, refraction, dispersion and energy flux etc.

\begin{table}[h]
\caption{The octonion multiplication table.}
\label{tab:table3}
\centering
\begin{tabular}{ccccccccc}
\hline\hline
$ $ & $1$ & $\emph{\textbf{i}}_1$  &
$\emph{\textbf{i}}_2$ & $\emph{\textbf{i}}_3$  &
$\emph{\textbf{I}}_0$ & $\emph{\textbf{I}}_1$
& $\emph{\textbf{I}}_2$  & $\emph{\textbf{I}}_3$  \\
\hline $1$ & $1$ & $\emph{\textbf{i}}_1$  & $\emph{\textbf{i}}_2$ &
$\emph{\textbf{i}}_3$  & $\emph{\textbf{I}}_0$  &
$\emph{\textbf{I}}_1$
& $\emph{\textbf{I}}_2$  & $\emph{\textbf{I}}_3$  \\
$\emph{\textbf{i}}_1$ & $\emph{\textbf{i}}_1$ & $-1$ &
$\emph{\textbf{i}}_3$  & $-\emph{\textbf{i}}_2$ &
$\emph{\textbf{I}}_1$
& $-\emph{\textbf{I}}_0$ & $-\emph{\textbf{I}}_3$ & $\emph{\textbf{I}}_2$  \\
$\emph{\textbf{i}}_2$ & $\emph{\textbf{i}}_2$ &
$-\emph{\textbf{i}}_3$ & $-1$ & $\emph{\textbf{i}}_1$  &
$\emph{\textbf{I}}_2$  & $\emph{\textbf{I}}_3$
& $-\emph{\textbf{I}}_0$ & $-\emph{\textbf{I}}_1$ \\
$\emph{\textbf{i}}_3$ & $\emph{\textbf{i}}_3$ &
$\emph{\textbf{i}}_2$ & $-\emph{\textbf{i}}_1$ & $-1$ &
$\emph{\textbf{I}}_3$  & $-\emph{\textbf{I}}_2$
& $\emph{\textbf{I}}_1$  & $-\emph{\textbf{I}}_0$ \\
\hline
$\emph{\textbf{I}}_0$ & $\emph{\textbf{I}}_0$ &
$-\emph{\textbf{I}}_1$ & $-\emph{\textbf{I}}_2$ &
$-\emph{\textbf{I}}_3$ & $-1$ & $\emph{\textbf{i}}_1$
& $\emph{\textbf{i}}_2$  & $\emph{\textbf{i}}_3$  \\
$\emph{\textbf{I}}_1$ & $\emph{\textbf{I}}_1$ &
$\emph{\textbf{I}}_0$ & $-\emph{\textbf{I}}_3$ &
$\emph{\textbf{I}}_2$  & $-\emph{\textbf{i}}_1$
& $-1$ & $-\emph{\textbf{i}}_3$ & $\emph{\textbf{i}}_2$  \\
$\emph{\textbf{I}}_2$ & $\emph{\textbf{I}}_2$ &
$\emph{\textbf{I}}_3$ & $\emph{\textbf{I}}_0$  &
$-\emph{\textbf{I}}_1$ & $-\emph{\textbf{i}}_2$
& $\emph{\textbf{i}}_3$  & $-1$ & $-\emph{\textbf{i}}_1$ \\
$\emph{\textbf{I}}_3$ & $\emph{\textbf{I}}_3$ &
$-\emph{\textbf{I}}_2$ & $\emph{\textbf{I}}_1$  &
$\emph{\textbf{I}}_0$  & $-\emph{\textbf{i}}_3$
& $-\emph{\textbf{i}}_2$ & $\emph{\textbf{i}}_1$  & $-1$ \\
\hline\hline
\end{tabular}
\end{table}

\psection{Wave equation for field strength in electromagnetic field}

The wave feature of gravitational field and electromagnetic field can be described simultaneously by the octonion space, which consists of two quaternion spaces.

In the quaternion space for the electromagnetic field, the basis vector is $\mathbb{E}_e$ = ($\emph{\textbf{I}}_0$, $\emph{\textbf{I}}_1$, $\emph{\textbf{I}}_2$, $\emph{\textbf{I}}_3$), the radius vector is $\mathbb{R}_e = \Sigma (R_i \emph{\textbf{I}}_i)$, and the velocity is $\mathbb{V}_e = \Sigma (V_i \emph{\textbf{I}}_i)$. The $\mathbb{E}_e$ is independent of the $\mathbb{E}_g$, with $\mathbb{E}_e = \mathbb{E}_g \circ \emph{\textbf{I}}_0$ and $\emph{\textbf{I}}_0 \circ \emph{\textbf{I}}_0 = -1 $. $V_0$ is the speed of electromagnetic intermediate boson.

These two quaternion spaces can be combined together to become an octonion space \cite{gogberashvili}, with the octonion basis vector $\mathbb{E} = (1, \emph{\textbf{i}}_1, \emph{\textbf{i}}_2, \emph{\textbf{i}}_3, \emph{\textbf{I}}_0, \emph{\textbf{I}}_1, \emph{\textbf{I}}_2, \emph{\textbf{I}}_3)$. The radius vector in the octonion space is $\mathbb{R} = \Sigma ( r_i \emph{\textbf{i}}_i + R_i \emph{\textbf{I}}_i ) $, and that the octonion velocity is $\mathbb{V} = \Sigma ( v_i \emph{\textbf{i}}_i + V_i \emph{\textbf{I}}_i) $ . Herein the symbol $\circ$ denotes the octonion multiplication.

The potential of gravitational field and electromagnetic field are $\mathbb{A}_g = \Sigma ( a_i \emph{\textbf{i}}_i)$ and $\mathbb{A}_e = \Sigma ( A_i \emph{\textbf{I}}_i)$ respectively. They are combined together to become the field potential, $\mathbb{A} = \mathbb{A}_g + k_{eg} \mathbb{A}_e$ , in the octonion space.
The field strength $\mathbb{B}$ consists of the gravitational strength $\mathbb{B}_g$ and electromagnetic strength $\mathbb{B}_e$ . That is, $\mathbb{B} = \lozenge \circ \mathbb{A} = \mathbb{B}_g + k_{eg} \mathbb{B}_e$ , with the $k_{eg}$ being a coefficient. The strength $\mathbb{B}_e$ covers two parts, $\textbf{E}/V_0 = ( \partial_0 \textbf{A} + \nabla A_0 ) \circ \emph{\textbf{I}}_0$ , and $\textbf{B} = ( \nabla \times \textbf{A} ) \circ \emph{\textbf{I}}_0$ . The gauge equations are $b_0 = 0$ and $B_0 = 0$. Herein $B_0 = \partial_0 A_0 + \nabla \cdot \textbf{A}$, with $\textbf{A} = \Sigma (A_j \emph{\textbf{i}}_j)$.

Comparing with the classical electromagnetic theory, the electromagnetic source is the current density $\mathbb{S}_e = q \mathbb{V}_e$ , while the gravitational source is the linear momentum density. And the source $\mathbb{S}$ was devised to describe consistently the field source of electromagnetism and gravitation.
\begin{eqnarray}
(\mathbb{B} / v_0 + \lozenge)^* \circ \mathbb{B} = - \mu \mathbb{S} = - (\mu_g \mathbb{S}_g + k_{eg} \mu_e \mathbb{S}_e)~,
\end{eqnarray}
where $\mu$ is a coefficient, and $\mu_e$ is the electromagnetic constant, with $k_{eg}^2 = \mu_g / \mu_e$ . $q$ is the electric charge, and $*$ denotes the conjugate of octonion.

According to the basis vectors, the above can be decomposed further as follows,
\begin{eqnarray}
\mathbb{B}^* \circ \mathbb{B} / v_0 + \lozenge^* \circ \mathbb{B}_g = - \mu_g \mathbb{S}_g~, ~~\lozenge^* \circ \mathbb{B}_e = - \mu_e \mathbb{S}_e~,
\end{eqnarray}
where the former equation is similar to Eq.(1) and is suitable for the gravitational field, while the latter equation is for the electromagnetic field.

Further, the latter equation in the above can be rewritten as follows
\begin{eqnarray}
& \nabla \cdot \textbf{B} = 0~, ~ \partial_0 \textbf{B} + \nabla^* \times \textbf{E} / V_0 = 0~,
\\
& \nabla^* \cdot \textbf{E} = - (q / \varepsilon_e) \emph{\textbf{I}}_0~, ~ \partial_0 \textbf{E} / V_0 + \nabla^* \times \textbf{B} = - \mu_e \textbf{S}~,
\end{eqnarray}
where the coefficient $\varepsilon_e = 1/(\mu_e V_0^2)$. $\textbf{S} = \Sigma (S_j \emph{\textbf{I}}_j) $, $S_j = q V_j $. $\textbf{S}_0 = S_0 \emph{\textbf{I}}_0$, $S_0 = q V_0$.

By contrast with Maxwell equation in terms of vector terminology, it is found that the above is the same as that in the conventional electromagnetic theory respectively, except for the direction of displacement current. And the gauge equation $B_0 = 0$ is different.

\begin{table}[h]
\caption{The quaternion, Squaternion, and octonion with the complex number.}
\label{tab:table4}
\centering
\begin{tabular}{llll}
\hline\hline
$subspace$
&   $meaning$               &  $coordinate$        &   $basis~vector$    \\
\hline
quaternion
&  quaternion               &  real number         &  $\emph{\textbf{i}}_0, \emph{\textbf{i}}_1, \emph{\textbf{i}}_2,
                                                       \emph{\textbf{i}}_3$ \\
&  bi-quaternion            &  imaginary number    &  $\emph{\textbf{i}}_0, \emph{\textbf{i}}_1, \emph{\textbf{i}}_2,
                                                       \emph{\textbf{i}}_3$ \\
&  complex quaternion       &  complex number      &  $\emph{\textbf{i}}_0, \emph{\textbf{i}}_1, \emph{\textbf{i}}_2,
                                                       \emph{\textbf{i}}_3$ \\
Squaternion
&  Squaternion              &  real number         &  $\emph{\textbf{I}}_0, \emph{\textbf{I}}_1, \emph{\textbf{I}}_2,
                                                       \emph{\textbf{I}}_3$ \\
&  bi-Squaternion           &  imaginary number    &  $\emph{\textbf{I}}_0, \emph{\textbf{I}}_1, \emph{\textbf{I}}_2,
                                                       \emph{\textbf{I}}_3$ \\
&  complex Squaternion      &  complex number      &  $\emph{\textbf{I}}_0, \emph{\textbf{I}}_1, \emph{\textbf{I}}_2,
                                                       \emph{\textbf{I}}_3$ \\
octonion
&  octonion                 &  real number         &  $\emph{\textbf{i}}_0, \emph{\textbf{i}}_1, \emph{\textbf{i}}_2,
                                                       \emph{\textbf{i}}_3, \emph{\textbf{I}}_0, \emph{\textbf{I}}_1,
                                                       \emph{\textbf{I}}_2, \emph{\textbf{I}}_3 $ \\
&  bi-octonion              &  imaginary number    &  $\emph{\textbf{i}}_0, \emph{\textbf{i}}_1, \emph{\textbf{i}}_2,
                                                       \emph{\textbf{i}}_3, \emph{\textbf{I}}_0, \emph{\textbf{I}}_1,
                                                       \emph{\textbf{I}}_2, \emph{\textbf{I}}_3 $ \\
&  complex octonion         &  complex number      &  $\emph{\textbf{i}}_0, \emph{\textbf{i}}_1, \emph{\textbf{i}}_2,
                                                       \emph{\textbf{i}}_3, \emph{\textbf{I}}_0, \emph{\textbf{I}}_1,
                                                       \emph{\textbf{I}}_2, \emph{\textbf{I}}_3 $ \\
\hline\hline
\end{tabular}
\end{table}

\psubsection{Octonion exponential function}

Similarly the octonion $\mathbb{O}$ is called as the bi-octonion when the coordinate is the imaginary number. Meanwhile the octonion $\mathbb{O}$ is the complex octonion with the complex number coordinate.

The octonion $\mathbb{O}$ is separated into two parts, the quaternion $\mathbb{O}_1$ and Squaternion $\mathbb{O}_2$ ,
\begin{eqnarray}
\mathbb{O} = \mathbb{O}_1 + \mathbb{O}_2 ~,~\mathbb{O}_2 = \emph{\textbf{I}}_0 \circ \mathbb{O}_3~,
\end{eqnarray}
where $\mathbb{O}_3$ is the quaternion. Squaternion means the second quaternion or step-quaternion.

In the octonion space, defining the complex Squaternion as
\begin{eqnarray}
exp \left\{ \textbf{I} ( a + i b ) \right\} = \emph{\textbf{I}}_0 \circ \left\{ cos ( a + i b ) + \textbf{i}' sin ( a + i b ) \right\} = \emph{\textbf{I}}_0 cos ( a + i b ) + \textbf{I} sin ( a + i b )~,
\end{eqnarray}
where $a$ and $b$ are all real. The $i$ is an imaginary unit, with $i^2 = -1$ . The quaternion unit $\textbf{i}'$ is one function of $\emph{\textbf{i}}_j$ , with $ \textbf{i}' \circ \textbf{i}' = -1$ . Meanwhile the unit $\textbf{I} = \emph{\textbf{I}}_0 \circ \textbf{i}'$ , with $ \textbf{I} \circ \textbf{I} = -1$ .

When $b = 0$, the above reduces to the Squaternion, $ exp ( \textbf{I} a ) = \emph{\textbf{I}}_0 cos ( a ) + \textbf{I} sin ( a )$ . Similar to Hamiltonian bi-quaternion, when $a = 0$, Eq.(29) reduces to the bi-Squaternion,
\begin{eqnarray}
exp ( \textbf{I} i b ) = \emph{\textbf{I}}_0 cos ( i b ) + \textbf{I} sin ( i b ) ~,
\end{eqnarray}
with
\begin{eqnarray}
exp ( - \textbf{I} i b ) = \emph{\textbf{I}}_0 cos ( i b ) - \textbf{I} sin ( i b ) ~.
\end{eqnarray}

The bi-Squaternion $exp(-i \textbf{I} \alpha)$ satisfies,
\begin{eqnarray}
& exp(-i \textbf{I} \alpha) \circ exp(i \textbf{I} \alpha) = \emph{\textbf{I}}_0 \circ exp(i \textbf{I} 2\alpha)~, ~ exp(i \textbf{I} \alpha) \circ exp(i \textbf{I} \alpha) =  -1~,
\\
& \partial_j exp(-i \textbf{I} \alpha) = i \textbf{I} K_j \circ exp(-i \textbf{I} \alpha) ~, ~ \partial_j^2 exp(-i \textbf{I} \alpha) = K_j^2 exp(-i \textbf{I} \alpha) ~, \end{eqnarray}
where $\alpha = - \Sigma (K_j r_j)$, with $K_j$ being the coefficient.

In case of neglecting the direction of $\emph{\textbf{I}}_0$ , the imaginary unit $i$ will replace the unit $\emph{\textbf{I}}_0$ , and then Eq.(29) reduces to the quaternion exponential function similar to Eq.(4).

\begin{table}[h]
\caption{The operator and multiplication of the physical quantity in the octonion space.}
\label{tab:table4}
\centering
\begin{tabular}{ll}
\hline\hline
definition                  &   meaning                                             \\
\hline
$\nabla \cdot \textbf{S}$   &  $-(\partial_1 S_1 + \partial_2 S_2 + \partial_3 S_3) \emph{\textbf{I}}_0 $  \\
$\nabla \times \textbf{S}$  &  $-\emph{\textbf{I}}_1 ( \partial_2
                                 S_3 - \partial_3 S_2 ) - \emph{\textbf{I}}_2 ( \partial_3 S_1
                                 - \partial_1 S_3 ) - \emph{\textbf{I}}_3 ( \partial_1 S_2
                                 - \partial_2 S_1 )$                                 \\
$\nabla \textbf{S}_0$       &  $\emph{\textbf{I}}_1 \partial_1 S_0
                                 + \emph{\textbf{I}}_2 \partial_2 S_0
                                 + \emph{\textbf{I}}_3 \partial_3 S_0  $             \\
$\partial_0 \textbf{S}$     &  $\emph{\textbf{I}}_1 \partial_0 S_1
                                 + \emph{\textbf{I}}_2 \partial_0 S_2
                                 + \emph{\textbf{I}}_3 \partial_0 S_3  $             \\
\hline\hline
\end{tabular}
\end{table}

\psubsection{Wave equation}

Studying the wave transmission feature in electromagnetic fields needs to solve Maxwell equation. In a vacuum far away from the electromagnetic sources, there does not exist free electric charge density $S_0$ and current density $\textbf{S}$, and then field equation can be reduced to,
\begin{eqnarray}
\lozenge^* \circ \mathbb{B}_e = 0~.
\end{eqnarray}

Applying the operator $\lozenge$ to the above, we have Laplace equation,
\begin{eqnarray}
\lozenge \circ (\lozenge^* \circ \mathbb{B}_e) = 0~, ~or ~ (\partial_0^2 + \partial_1^2 + \partial_2^2 + \partial_3^2) \mathbb{B}_e = 0~.
\end{eqnarray}

The Laplace equation is the wave equation in the octonion space. This wave equation can be obtain from Eq.(34) with the conjugate operator $\lozenge^*$ and the field equation also.

Proceeding with the operator $\partial_0$ and $\nabla$, we can obtain the wave equation about the components of field strength from Maxwell equation Eqs.(26) and (27) directly,
\begin{eqnarray}
 (\partial_0^2 + \partial_1^2 + \partial_2^2 + \partial_3^2) \textbf{E} = 0~,
~ (\partial_0^2 + \partial_1^2 + \partial_2^2 + \partial_3^2) \textbf{B} = 0~.
\end{eqnarray}

The above means that the Eq.(35) can be decomposed into two equations in the above. And two strength components of electromagnetic field, $\textbf{E}$ and $\textbf{B}$, are both possessed of wave features, and can be detected at present.

\psubsection{Transverse wave}

In a similar way, for the electromagnetic wave with the angular frequency $\omega$, the field strength should be a harmonic function, $cos \omega t$, and can be chosen as the function $exp(-i \omega t)$ .

The electromagnetic strength $\textbf{E}$ and $\textbf{B}$ can be written as follows,
\begin{eqnarray}
\textbf{E} = \textbf{E}(r) exp(- i \omega t ) ~,~ \textbf{B} = \textbf{B}(r) exp(- i \omega t ) ~,
\end{eqnarray}
substituting the above in Eq.(36),
\begin{eqnarray}
\left\{ - ( \omega / v_0 )^2 + \Sigma \partial_j^2 \right\} \textbf{E}(r) = 0~, ~ \left\{ - ( \omega / v_0 )^2 + \Sigma \partial_j^2 \right\} \textbf{B}(r) = 0~.
\end{eqnarray}

From the Maxwell equations, $\partial_0 \textbf{B} + \nabla^* \times \textbf{E} / V_0 = 0$ and $\partial_0 \textbf{E} / V_0 + \nabla^* \times \textbf{B} = 0$, we find that the amplitude of electromagnetic wave will be increased steadily or decreased continuously, although this variation may be quite slow. So the field strength is the hyperbolic cosine $cos ( i \alpha)$, and can be replaced by $exp(-i \textbf{I} \alpha)$.

And then, the field strength $\textbf{E}(r)$ and $\textbf{B}(r)$ are
\begin{eqnarray}
\textbf{E}(r) = \textbf{E}_0 \circ exp(-i \textbf{I} \alpha) ~,
~ \textbf{B}(r) = \textbf{B}_0 \circ exp(-i \textbf{I} \alpha) ~,
\end{eqnarray}
where $\textbf{E}_0$ and $\textbf{B}_0$ both are the constant vectors in the octonion space.

Further, substituting Eq.(39) in Eq.(38), we have the result,
\begin{eqnarray}
- (\omega / v_0 )^2 + \Sigma K_j^2 = 0~.
\end{eqnarray}

In the electromagnetic field, we will find that the electromagnetic waves are the transverse waves in a vacuum. From the field equations
\begin{eqnarray}
\nabla \cdot \textbf{E} = 0~,~\nabla \cdot \textbf{B} = 0~,
\end{eqnarray}
we obtain
\begin{eqnarray}
\textbf{K} \cdot \textbf{E}_0 = 0~, ~\textbf{K} \cdot \textbf{E}'_0 = 0~, ~\textbf{K} \cdot \textbf{B}_0 = 0~, ~\textbf{K} \cdot \textbf{B}'_0 = 0~,
\end{eqnarray}
where $\textbf{E}'_0 = \textbf{E}_0 \circ \textbf{I}$ , $\textbf{B}'_0  =  \textbf{B}_0 \circ \textbf{I}$ . The wave vector $\textbf{K} = \Sigma (\emph{\textbf{I}}_j K_j)$ .

The above states that the electromagnetic waves, $\textbf{E}_0$ and $\textbf{B}_0$ , belong to the transverse waves. So do the two new wave components, $\textbf{E}'_0$ and $\textbf{B}'_0$ . Moreover, the amplitudes of wave components, $\textbf{E}'_0$ and $\textbf{B}'_0$, are equal to that of $\textbf{E}_0$ and $\textbf{B}_0$ respectively.

In the electromagnetic field, there exist some relationships among the wave components, $\textbf{E}_0$ and $\textbf{B}_0$ , with two new kinds of wave components, $\textbf{E}'_0$ and $\textbf{B}'_0$ . The field equations,
\begin{eqnarray}
\nabla \times \textbf{E} = k_v \partial \textbf{B} / \partial t~,~\nabla \cdot \textbf{E} = 0~,
\end{eqnarray}
can be rewritten as,
\begin{eqnarray}
\nabla \circ \textbf{E} = k_v \partial \textbf{B} / \partial t~.
\end{eqnarray}
where the coefficient $k_v = V_0 / v_0$, with $k_v \approx 1$ in general.

Expanding the $exp(-i \textbf{I} \alpha)$, the above implies that
\begin{eqnarray}
[ \textbf{K} \circ \textbf{E}'_0 + \omega k_v \textbf{B}_0 ] cosh (\alpha )
+ [ \textbf{K} \circ \textbf{E}_0 - \omega k_v \textbf{B}'_0 ] sinh (\alpha ) = 0~,
\end{eqnarray}
considering Eq.(42) and the $\alpha$ may be any value, and then
\begin{eqnarray}
\textbf{K} \times \textbf{E}'_0 + k_v \omega \textbf{B}_0 = 0~,
~\textbf{K} \times \textbf{E}_0 - k_v \omega \textbf{B}'_0 = 0~.
\end{eqnarray}

In the same way, from the field equations
\begin{eqnarray}
v_0 V_0 \nabla \times \textbf{B} = \partial \textbf{E} / \partial t~,~\nabla \cdot \textbf{B} = 0~,
\end{eqnarray}
we have
\begin{eqnarray}
\textbf{K} \times \textbf{B}'_0 + \omega (\textbf{E}_0 / V_0) / v_0 = 0~,
~\textbf{K} \times \textbf{B}_0 - \omega (\textbf{E}'_0 / V_0) / v_0 = 0~.
\end{eqnarray}

In the above, the $\textbf{K}$ and $\textbf{E}_0$ will yield a new wave component $\textbf{B}'_0$ , while the $\textbf{K}$ and $\textbf{B}_0$ produce a new wave component $\textbf{E}'_0$ . Associating with the $\textbf{K} \times \textbf{B}_0$ is only the $\textbf{E}'_0$, and with the $\textbf{K} \times \textbf{E}_0$ is the $\textbf{B}'_0$ . This implies that it is impossible to determine simultaneously the $\textbf{E}_0$ and $\textbf{B}_0$ via the $\textbf{K}$. However, we can measure the $\textbf{E}_0$ and $\textbf{B}'_0$ synchronously, or the $\textbf{E}'_0$ and $\textbf{B}_0$ at the same time. There are many electric substances being able to slow down the speed of electromagnetic waves, and the field strength will possess the wave features, including the reflection and refraction etc.

\psubsection{Reflection and refraction}

In the electromagnetic field, by means of above features and the boundary condition, we can achieve the law of reflection, law of refraction, Fresnel formula, and the total internal reflection, etc.

The electromagnetic wave transmits from the first medium to the second one, while there is one interface between these two media. In this interface, the $\textbf{t}$ is the vector unit along to a tangent direction. The tangent components $\textbf{E}_t$ of $\textbf{E}$, and $\textbf{B}'_t$ of $\textbf{B}'$ will be continuous respectively, for there is not electric current in the interface. There are the incident wave $\mathbb{B}_e^a$ and reflective wave $\mathbb{B}_e^b$ in the first medium, and the refractive wave $\mathbb{B}_e^c$ in the second medium, with $\textbf{H}' = \textbf{B}' / \mu$ . Therefore the boundary condition in the interface are
\begin{eqnarray}
E^a_t + E^b_t = E^c_t~, ~ (B'^a_t + B'^b_t)/\mu_1 = B'^c_t/\mu_2~,
\end{eqnarray}
where $\textbf{E}^f_t = \textbf{t} E^f_t$ , $\textbf{B}'^f_t = \textbf{t} B'^f_t$ , with $f = a, b, c$ . The $f = a$ stands for the incident wave, meanwhile $f = b$ and $f = c$ for the reflective and refractive waves respectively. $\mu_1$ and $\varepsilon_1$ are the permeability and permittivity for the first medium respectively. While $\mu_2$ and $\varepsilon_2$ for the second medium.

In the Cartesian coordinate system $oxyz$, setting the plane $z = 0$ on the interface, and the incident wave on the plane $y = 0$ , the $z$-axis points to the second medium. In the incident plane, there is $y = 0$ , and then $K^a_y = 0$ . In the interface, there exists $z = 0$, therefore
\begin{eqnarray}
&& E^a_{0t} exp(i \textbf{I} x K^a_x) exp(i \omega^a t) + E^b_{0t} exp[i \textbf{I} (x K^b_x + y K^b_y) ] exp(i \omega^b t) \nonumber
\\
= && E^c_{0t} exp[i \textbf{I} (x K^c_x + y K^c_y)] exp(i \omega^c t)~,~
\\
&& B'^a_{0t} exp(i \textbf{I} x K^a_x) exp(i \omega^a t)/\mu_1 + B'^b_{0t} exp[i \textbf{I} (x K^b_x + y K^b_y) ] exp(i \omega^b t)/\mu_1  \nonumber
\\
= && B'^c_{0t} exp[i \textbf{I} (x K^c_x + y K^c_y)] exp(i \omega^c t) /\mu_2~,
\end{eqnarray}
where the wave vector $\textbf{K}^f = \emph{\textbf{I}}_1 K^f_x + \emph{\textbf{I}}_2 K^f_y + \emph{\textbf{I}}_3 K^f_z$ , the wave number $K^f = \omega^f / V^f_0 $ .

Equating exponential parts for each coordinate and time,
\begin{eqnarray}
x K^a_x  = x K^b_x + y K^b_y =  x K^c_x + y K^c_y ~,~ \omega^a t =  \omega^b t = \omega^c t~,
\end{eqnarray}
and then
\begin{eqnarray}
K^a_x = K^b_x = K^c_x~,~ \omega^a = \omega^b = \omega^c~,~K^b_y = K^c_y~.
\end{eqnarray}

The above means that the angular frequencies of reflective wave and of refractive wave both are equal to that of incident wave. The refracted ray and reflected ray both are in the plane $y = 0$ , that is these two rays and incident ray stand in same one plane. In the result, we will neglect the subscript ($_t$) in the following context.

According to the above results and definition of refractive index, we get
\begin{eqnarray}
K^a_x = K^a sin \theta^a~,~K^b_x = K^b sin \theta^b~,~K^c_x = K^c sin \theta^c~,
\end{eqnarray}
therefore
\begin{eqnarray}
\theta^a = \theta^b~,~sin \theta^a / sin \theta^c = n_{21}~,
\end{eqnarray}
where $\theta^a$, $\theta^b$, and $\theta^c$ are the angle of incidence, angle of reflection, and angle of refraction respectively. $n_{21}$ is the refractivity of the second medium relative to the first one.

The above results are so-called law of reflection and law of refraction for the light.

\psubsection{Fresnel formula}

In the electromagnetic field, Fresnel formula can be derived from the relationships among the amplitudes and angles of incident wave, reflective wave, and refractive wave.

The boundary condition of two media interface can be reduced to
\begin{eqnarray}
E^a_0 + E^b_0 = E^c_0 ~,~ (B'^a_0  + B'^b_0) /\mu_1 = B'^c_0 /\mu_2~.
\end{eqnarray}

For the high-frequency vibration of electromagnetic waves, the $\mu$ of each media is almost equal to $\mu_0$ . And then the refractive index is
\begin{eqnarray}
n_{21} = sin \theta^a / sin \theta^c = \sqrt{ \varepsilon_2 / \varepsilon_1 }~.
\end{eqnarray}

When the $\textbf{E}^a$ is perpendicular to the incident plane, Eq.(46) says that
\begin{eqnarray}
B'^a_0 = \sqrt{\varepsilon_1 \mu_0} E^a_0 ~,~B'^b_0 = \sqrt{\varepsilon_1 \mu_0} E^b_0 ~,~B'^c_0 = \sqrt{\varepsilon_2 \mu_0} E^c_0 ~,
\end{eqnarray}
combining this with Eq.(56), gives
\begin{eqnarray}
E^a_0 + E^b_0 = E^c_0~,~ - \sqrt{\varepsilon_1} E^a_0 cos \theta^a + \sqrt{\varepsilon_1} E^b_0 cos \theta^b = - \sqrt{\varepsilon_2} E^c_0 cos \theta^c~,
\end{eqnarray}
and then
\begin{eqnarray}
E^b_0 / E^a_0 = - sin ( \theta^a - \theta^c ) / sin ( \theta^a + \theta^c )~,~E^c_0 / E^a_0 = 2 cos \theta^a sin \theta^c / sin ( \theta^a + \theta^c )~.
\end{eqnarray}

In the case of the $\textbf{E}^a$ is parallel to the incident plane, we have
\begin{eqnarray}
E^a_0 cos \theta^a - E^b_0 cos \theta^b = E^c_0 cos \theta^c~, ~ \sqrt{\varepsilon_1} E^a_0 + \sqrt{\varepsilon_1} E^b_0 = \sqrt{\varepsilon_2} E^c_0~,
\end{eqnarray}
substituting Eq.(57) in the above, therefore
\begin{eqnarray}
E^b_0 / E^a_0 = tg ( \theta^a - \theta^c ) / tg ( \theta^a + \theta^c )~,~E^c_0 / E^a_0 = 2 cos \theta^a sin \theta^c / sin ( \theta^a + \theta^c ) cos ( \theta^a + \theta^c )~.
\end{eqnarray}

In the electromagnetic field, from Eqs.(46), (49), and (57), we obtain Fresnel formula, Eqs.(60) and (62), related to the $\textbf{E}$ . On the other hand, we can find similar inferences about the $\textbf{E}'$ , and some other results including Brewster's angle and the half-wave loss etc.

\psubsection{Total internal reflection}

When the light passes from a denser medium to the interface of the less dense medium, $ \varepsilon_1 > \varepsilon_2 $ , $ n_{21} < 1 $ , and $ \theta^c > \theta^a $ . Increasing the incident angle $\theta^a$ to the critical angle, the $\theta^c$ will achieve $\pi / 2$ . Meanwhile the refracted ray sweeps along the interface, and $ n_{21} = sin \theta^a = \sqrt{\varepsilon_2 / \varepsilon_1} $ .

Enlarging the incident angle continuously, we will find that $sin \theta^a > n_{21}$ , and then
\begin{eqnarray}
sin \theta^c = sin \theta^a / n_{21} > 1~, ~ cos \theta^c = i \sqrt{ (sin \theta^c)^2 - 1 }~,
\end{eqnarray}
that means the refractive angle $\theta^c$ becomes an imaginary number. At the moment, the $x$-component and $z$-component of refracted wave vector $\textbf{K}^c$ are respectively,
\begin{eqnarray}
K^c_x = K^c sin \theta^c > K^c~,~K^c_z = \pm i \sqrt{ (K^c_x)^2 - (K^c)^2}~.
\end{eqnarray}

Considering $ K^c_x = K^a_x $ , and $\Sigma (K^c_j r_j) = x K^a_x + z K^c_z$ , the electromagnetic strength will be
\begin{eqnarray}
\textbf{E}^c = \textbf{E}^c_0 \circ exp[ - z \textbf{I} \sqrt{ (K^c_x)^2 - (K^c)^2} ] \circ exp( - i \textbf{I} x K^a_x ) exp(i \omega t)~,
\\
\textbf{B}^c = \textbf{B}^c_0 \circ exp[ - z \textbf{I} \sqrt{ (K^c_x)^2 - (K^c)^2} ] \circ exp( - i \textbf{I} x K^a_x ) exp(i \omega t)~.
\end{eqnarray}

It is easy to find that the $exp[ - z \textbf{I} \sqrt{ (K^c_x)^2 - (K^c)^2} ]$ is an octonion quantity. The above implies that the electromagnetic transmitting along the $z$-axis will be attenuated quickly. Eventually the incident wave is totally reflected inside. And that the amplitude and energy flux density of reflective wave are equal to that of incident wave respectively.

\begin{table}[h]
\caption{Some wave equations in the electromagnetic and gravitational fields.} \label{tab:table1}
\centering
\begin{tabular}{llllc}
\hline\hline
$definition$                                          & $meaning$                   & $condition$       & $wave~equation$    & $separability$ \\
\hline
$\mathbb{X}$                                          & field quantity \\
$\mathbb{A} = \lozenge \circ \mathbb{X}$              & field potential             & $\mathbb{A}=0$    & $\lozenge^* \circ (\lozenge \circ \mathbb{X}) = 0$   &  no  \\
$\mathbb{B} = \lozenge \circ \mathbb{A}$              & field strength              & $\mathbb{B}=0$    & $\lozenge^* \circ (\lozenge \circ \mathbb{A}) = 0$ &  no\\
$\mu \mathbb{S} = - ( \mathbb{B} /v_0 + \lozenge )^* \circ \mathbb{B}$              & field source      & $\mu \mathbb{S} + \mathbb{B}^*  \circ \mathbb{B} /v_0 = 0$    & $\lozenge \circ (\lozenge^* \circ \mathbb{B}) = 0$ & yes  \\
$\mathbb{P} = \mu \mathbb{S} / \mu_g$                 & linear momentum \\
$\mathbb{L} = (\mathbb{R} + k_{rx} \mathbb{X}) \circ \mathbb{P}$                    & angular momentum \\
$\mathbb{W} = v_0 ( \mathbb{B} /v_0 + \lozenge ) \circ \mathbb{L}$                  & energy and torque            & $\mathbb{W} - \mathbb{B} \circ \mathbb{L} = 0 $    & $\lozenge^* \circ (\lozenge \circ \mathbb{L}) = 0$   &  no  \\
$\mathbb{N} = v_0 ( \mathbb{B} /v_0 + \lozenge )^* \circ \mathbb{W}$                & power and force            & $\mathbb{N} - \mathbb{B}^* \circ \mathbb{W} = 0 $    & $\lozenge \circ (\lozenge^* \circ \mathbb{W}) = 0$   &  yes  \\
$\mathbb{C} = v_0 ( \mathbb{B} /v_0 + \lozenge )^* \circ \mathbb{N}$                & physical quantity             & $\mathbb{C} - \mathbb{B}^* \circ \mathbb{N} = 0 $    & $\lozenge \circ (\lozenge^* \circ \mathbb{N}) = 0$   &  yes  \\
\hline\hline
\end{tabular}
\end{table}

\psubsection{Other wave equations}

In the octonion space, the field potential $\mathbb{A}$ is defined from the field quantity $\mathbb{X}$, which can be considered as the integral of field potential. And we can define the angular momentum $\mathbb{L}$, energy-torque  quantity $\mathbb{W}$, power-force quantity $\mathbb{N}$, and physical quantity $\mathbb{C}$ etc, from the radius vector $\mathbb{R}$, field quantity $\mathbb{X}$, and linear momentum $\mathbb{P}$.

Similarly there are some other kinds of wave equations in the electromagnetic and gravitational fields, including that for the field potential and angular momentum etc.

\psection{Torque wave and force wave in gravitational field}

To incorporate various kinds of energies within a single definition, the angular momentum and energy will both be extended to apply within gravitational fields. The angular momentum density $\mathbb{L}_g = l_0 + \Sigma (l_j \emph{\textbf{i}}_j )$ is defined from the linear momentum density $\mathbb{P}_g = \mu_g \mathbb{S}_g / \mu_g$ in Eq.(1) and the radius vector $\mathbb{R}_g$ . And it can be rewritten as follows,
\begin{eqnarray}
\mathbb{L}_g = (\mathbb{R}_g + k_{rx} \mathbb{X}_g ) \circ
\mathbb{P}_g~,
\end{eqnarray}
where $l_0 = (r_0 + k_{rx} x_0) p_0 + (\textbf{r} + k_{rx} \textbf{x}) \cdot \textbf{p}~, \textbf{l} = (r_0 + k_{rx} x_0) \textbf{p} + p_0 (\textbf{r} + k_{rx} \textbf{x}) + (\textbf{r} + k_{rx} \textbf{x}) \times \textbf{p} $, and $\textbf{l} = \Sigma (l_j \emph{\textbf{i}}_j )$. The quaternion quantity $\mathbb{X}_g = \Sigma (x_i \emph{\textbf{i}}_i)$ is similar to Hertz vector in the electrodynamics theory, with $k_{rx}\mathbb{X}_g \ll \mathbb{R}_g$ and $k_{rx} = 1$ . The derivation of $\mathbb{X}_g$ will yield the gravitational potential. The $a_0 = v_0 (\partial_0 x_0 + \nabla \cdot \textbf{x})$ and $\textbf{a} = v_0 (\partial_0 \textbf{x} + \nabla x_0 + \nabla \times \textbf{x})$
are the scalar potential and vectorial potential of gravitational field respectively. $\textbf{x} = \Sigma (x_j \emph{\textbf{i}}_j )$.

We choose the following definition of energy to include various energies in the gravitational field. In quaternion space, the quaternion energy-torque density $\mathbb{W}_g = w_0 + \Sigma (w_j \emph{\textbf{i}}_j )$ is defined from the angular momentum density
$\mathbb{L}_g$ ,
\begin{eqnarray}
\mathbb{W}_g = v_0 ( \mathbb{B}_g/v_0 + \lozenge) \circ \mathbb{L}_g~,
\end{eqnarray}
where $w_0 = v_0 \partial_0 l_0 + v_0 \nabla \cdot \textbf{l} + (\textbf{g} / v_0 + \textbf{b}) \cdot \textbf{l}
$, and the $-w_0/2$ is the energy density, including the density of kinetic energy, potential energy, field energy, and work etc. While the $\textbf{w} = \Sigma (w_j \emph{\textbf{i}}_j ) $, with $\textbf{w} = v_0 \partial_0 \textbf{l} + v_0 \nabla l_0 + v_0 \nabla \times \textbf{l} + l_0 (\textbf{g} / v_0 + \textbf{b}) + (\textbf{g} / v_0 + \textbf{b}) \times \textbf{l}$. And the $-\textbf{w}/2$ is the torque density, including that caused by gravity and some other force terms.

The quaternion power-force density $\mathbb{N}_g = n_0 + \Sigma (n_j \emph{\textbf{i}}_j )$ is defined from the $\mathbb{W}_g$ ,
\begin{eqnarray}
\mathbb{N}_g = v_0 ( \mathbb{B}_g/v_0 + \lozenge)^* \circ \mathbb{W}_g~,
\end{eqnarray}
where $ n_0 =  v_0 \partial_0 w_0 + v_0 \nabla^* \cdot \textbf{w} + (\textbf{g}/v_0 + \textbf{b})^* \cdot \textbf{w} $, and $f_0 = -n_0/(2 v_0)$ is the power density. While $\textbf{n}  = \Sigma (n_j \emph{\textbf{i}}_j ) = v_0 \nabla^* w_0  + v_0 \partial_0 \textbf{w} + v_0 \nabla^* \times \textbf{w} + (\textbf{g}/v_0 + \textbf{b})^* \times \textbf{w} + w_0 (\textbf{g}/v_0 + \textbf{b})^*$. The force density $\textbf{f}$ in the gravitational field can be defined from $\textbf{n}$ , that is, $\textbf{f} = - \textbf{n} /(2 v_0)$ . The force density $\textbf{f}$ includes the density of gravity, inertial force, and energy gradient etc.

To deduce the wave equation for force, the physical quantity $\mathbb{C}_g = c_0 + \Sigma (c_j \emph{\textbf{i}}_j )$ can similarly be defined as follows,
\begin{eqnarray}
\mathbb{C}_g = v_0 ( \mathbb{B}_g/v_0 + \lozenge)^* \circ \mathbb{N}_g~,
\end{eqnarray}
where $ c_0 =  v_0 \partial_0 n_0 + v_0 \nabla^* \cdot \textbf{n} + (\textbf{g}/v_0 + \textbf{b})^* \cdot \textbf{n} $, while $\textbf{c}  = \Sigma (c_j \emph{\textbf{i}}_j ) = v_0 \nabla^* n_0  + v_0 \partial_0 \textbf{n} + v_0 \nabla^* \times \textbf{n} + (\textbf{g}/v_0 + \textbf{b})^* \times \textbf{n} + n_0 (\textbf{g}/v_0 + \textbf{b})^*$. The $\textbf{c}$ includes the derivative of acceleration (or jerk).

\psubsection{Torque wave equation}

Choosing the gauge equation, $w_0 = 0$, from Eq.(68) to simplify following calculation. The torque density $\textbf{w}$ can be separated into three components, $\textbf{w} = \textbf{w}_e / v_0 + \textbf{w}_b + \textbf{w}_s$ ,
\begin{eqnarray}
& \textbf{w}_e / v_0 = v_0 \partial_0 \textbf{l} + v_0 \nabla l_0~,~ \textbf{w}_b = v_0 \nabla \times \textbf{l}~,
\\
& \textbf{w}_s = l_0 (\textbf{g} / v_0 + \textbf{b}) + (\textbf{g} / v_0 + \textbf{b}) \times \textbf{l}~.
\end{eqnarray}

\psubsubsection{Case A}

Similar to Eqs.(2) and (3), in case of $w_0 = 0$ and $\textbf{w}_s = 0$, the definition equation of quaternion power-torque density, Eq.(69), can be decomposed into the Maxwell-like equations,
\begin{eqnarray}
& \nabla \cdot (\textbf{w}_b / v_0) = 0~, ~\partial_0 (\textbf{w}_b / v_0) + \nabla^* \times (\textbf{w}_e / v_0) / v_0 = 0 ~,
\\
& \nabla^* \cdot (\textbf{w}_e / v_0) / v_0 = n_0 / v_0~, ~ \partial_0 (\textbf{w}_e / v_0) / v_0 + \nabla^* \times (\textbf{w}_b  / v_0) = \textbf{n} / v_0 ~.
\end{eqnarray}

In the above, the $\textbf{w}_s = 0$ is equivalent to $\textbf{g} / v_0 + \textbf{b} = 0$ from Eq.(72) when $\mathbb{L} \neq 0$. If there exists further $\mathbb{N}_g = 0$ , the above or Eq.(69) can be reduced to,
\begin{eqnarray}
\lozenge^* \circ \mathbb{W}_g = 0~.
\end{eqnarray}

Applying the operator $\lozenge$ to the above, we get,
\begin{eqnarray}
(\partial_0^2 + \partial_1^2 + \partial_2^2 + \partial_3^2) \mathbb{W}_g = 0~.
\end{eqnarray}

The above is Laplace equation as well as the wave equation for the torque in the quaternion space. Proceeding with the operator $\partial_0$ and $\nabla$, we can obtain the wave equation about the torque components from Eqs.(73) and (74) directly,
\begin{eqnarray}
 (\partial_0^2 + \partial_1^2 + \partial_2^2 + \partial_3^2) \textbf{w}_e = 0~,
~ (\partial_0^2 + \partial_1^2 + \partial_2^2 + \partial_3^2) \textbf{w}_b = 0~.
\end{eqnarray}

The above means that two components of torque, $\textbf{w}_e$ and $\textbf{w}_b$, are both possessed of the wave features in the gravitational field. And these torque component is similar to the periodic torque in the vibrational mechanics partially, and may cause the vortex street in the fluid mechanics.

\begin{table}[h]
\caption{Maxwell-like equations for field potential and for angular momentum.}
\label{tab:table2}
\centering
\begin{tabular}{lll}
\hline\hline
& field potential                          &                angular momentum                          \\
\hline
field strength   &  $\textbf{g}/v_0 = \partial_0 \textbf{a} + \nabla a_0 $ & $\textbf{w}_e / v_0 = v_0 \partial_0 \textbf{l} + v_0 \nabla l_0$ \\
$$   &  $ \textbf{b} = \nabla \times \textbf{a}$ &  $\textbf{w}_b = v_0 \nabla \times \textbf{l}$ \\
gauge equation  &  $b_0 = 0$ &  $w_0 = 0$ \\
$$   &  $ \sim $ &  $\textbf{w}_s = 0$ \\
field equation   &  $- \mu \mathbb{S}_g = (\mathbb{B}_g / v_0 + \lozenge)^* \circ \mathbb{B}_g~~$ & $- \mu \mathbb{N}_g = ( \mathbb{B}_g/v_0 + \lozenge)^* \circ \mathbb{W}_g$  \\
$$   &  $\mu = \mu_g $ & $\mu = - 1/ v_0$ \\
Maxwell-like equations   &  $\nabla \cdot \textbf{b} = 0$  &  $\nabla \cdot (\textbf{w}_b / v_0) = 0$ \\
$$   &  $\partial_0 \textbf{b} + \nabla^* \times \textbf{g} / v_0 = 0$  &  $\partial_0 (\textbf{w}_b / v_0) + \nabla^* \times (\textbf{w}_e / v_0) / v_0 = 0$  \\
$$   &  $\nabla^* \cdot \textbf{g} = - \widehat{m} / \varepsilon_g$  & $\nabla^* \cdot (\textbf{w}_e / v_0) / v_0 = n_0 / v_0$  \\
$$   &  $\partial_0 \textbf{g} / v_0 + \nabla^* \times \textbf{b} = - \mu_g \textbf{s}$  & $\partial_0 (\textbf{w}_e / v_0) / v_0 + \nabla^* \times (\textbf{w}_b  / v_0) = \textbf{n} / v_0$ \\
\hline\hline
\end{tabular}
\end{table}

\psubsubsection{Case B}

In case of $w_0 = 0$ and $\textbf{w}_s \neq 0$, Eq.(69) can be decomposed as follows,
\begin{eqnarray}
& n_0 / v_0 = \nabla^* \cdot \textbf{w} + (\textbf{g}/v_0 + \textbf{b})^* \cdot \textbf{w} / v_0 ~,
\\
& \textbf{n} / v_0 = \partial_0 \textbf{w} + \nabla^* \times \textbf{w} + (\textbf{g}/v_0 + \textbf{b})^* \times \textbf{w} / v_0 ~.
\end{eqnarray}

In general the above can be separated into the equations more complex than Eqs.(73) and (74), and may possess more complicated wave features than which in Eqs.(76) and (77).

Specially there may exist following conditions,
\begin{eqnarray}
& \nabla^* \cdot ( \textbf{w}_s + \textbf{w}_b ) + (\textbf{g}/v_0 + \textbf{b})^* \cdot \textbf{w} / v_0 = 0 ~,
\\
& \partial_0 ( \textbf{w}_s + \textbf{w}_b ) + \nabla^* \times ( \textbf{w}_e / v_0 + \textbf{w}_s) + (\textbf{g}/v_0 + \textbf{b})^* \times \textbf{w} / v_0 = 0 ~,
\end{eqnarray}
and considering Eq.(73), the above conditions can be reduced to,
\begin{eqnarray}
& \nabla^* \cdot \textbf{w}_s  + (\textbf{g}/v_0 + \textbf{b})^* \cdot \textbf{w} / v_0 = 0 ~,
\\
& \partial_0 \textbf{w}_s + \nabla^* \times \textbf{w}_s + (\textbf{g}/v_0 + \textbf{b})^* \times \textbf{w} / v_0 = 0 ~.
\end{eqnarray}

Under the above conditions when $\textbf{w}_s \neq 0$, Eqs.(78) and (79) can be reduced to the Maxwell-like equations, which is same as Eqs.(73) and (74).

If there exists additionally $\mathbb{N}_g = 0$ , Eqs.(78) and (79) can further be simplified to,
\begin{eqnarray}
\lozenge^* \circ \mathbb{W}_g = 0~,
\end{eqnarray}
and then we have Eqs.(76) and (77), and some similar inferences about the torque components.

In Eq.(69), when there exists the condition relation $\mathbb{N}_g - \mathbb{B}_g^* \circ \mathbb{W}_g = 0$ , we should gain some more complex deductions than the above directly.

\psubsubsection{Transverse wave}

For the torque wave with angular frequency $\omega$, the torque wave should be a harmonic function $cos \omega t$ ,  and then should be chosen as the function $exp(-i \omega t)$ . The torque wave $\textbf{w}_e$ and $\textbf{w}_b$ can be written as follows,
\begin{eqnarray}
\textbf{w}_e = \textbf{w}_e (r) exp(- i \omega t ) ~,~ \textbf{w}_b = \textbf{w}_b (r) exp(- i \omega t ) ~,
\end{eqnarray}
combining this with Eq.(76), gives
\begin{eqnarray}
\left\{ - ( \omega / v_0 )^2 + \Sigma \partial_j^2 \right\} \textbf{w}_e (r) = 0~, ~ \left\{ - ( \omega / v_0 )^2 + \Sigma \partial_j^2 \right\} \textbf{w}_b (r) = 0~.
\end{eqnarray}

From the analysis of Eqs.(73) and (74), we find that the torque wave should be one hyperbolic cosine $cos ( i \beta)$ , and then should be chosen as the function $exp(-i \textbf{i} \beta)$ . The wave-like vector $\textbf{k}_w = \Sigma (\emph{\textbf{i}}_j k_{wj})$ , while the $ \beta = - \Sigma ( r_j k_{wj} ) $ , with $k_{wj}$ being the coefficient.

The torque wave $\textbf{w}_e (r)$ and $\textbf{w}_b (r)$ are
\begin{eqnarray}
\textbf{w}_e (r) = \textbf{w}_{e0} \circ exp(-i \textbf{i} \beta) ~,
~ \textbf{w}_b (r) = \textbf{w}_{b0} \circ exp(-i \textbf{i} \beta) ~,
\end{eqnarray}
where $\textbf{w}_{e0}$ and $\textbf{w}_{b0}$ both are constant vectors in the quaternion space.

And then we have the result from the above and Eq.(85) ,
\begin{eqnarray}
- (\omega / v_0 )^2 + \Sigma k_{wj}^2 = 0~.
\end{eqnarray}

In the gravitational field, we will find that torque waves are the transverse waves in a vacuum. Similar to Eqs.(16)$\sim$(23), there are following results from Eqs.(73) and (74),
\begin{eqnarray}
& \textbf{k}_w \cdot \textbf{w}_{e0} = 0~, ~\textbf{k}_w \cdot \textbf{w}'_{e0} = 0~, ~\textbf{k}_w \cdot \textbf{w}_{b0} = 0~, ~\textbf{k}_w \cdot \textbf{w}'_{b0} = 0~,\\
& \textbf{k}_w \times \textbf{w}'_{e0} + \omega \textbf{w}_{b0} = 0~,
~\textbf{k}_w \times \textbf{w}_{e0} - \omega \textbf{w}'_{b0} = 0~,\\
&\textbf{k}_w \times \textbf{w}'_{b0} + \omega (\textbf{w}_{e0} / v_0) / v_0 = 0~,
~\textbf{k}_w \times \textbf{w}_{b0} - \omega (\textbf{w}'_{e0} / v_0) / v_0 = 0~,
\end{eqnarray}
where $\textbf{w}'_{e0} = \textbf{w}_{e0} \circ \textbf{i}$ , $\textbf{w}'_{b0} = \textbf{w}_{b0} \circ \textbf{i}$ .

The above means that there are some relationships between $\mathbb{W}_g$ with $\mathbb{W}_g \circ \textbf{i}$ . The $\textbf{k}_w$ and $\textbf{w}_{e0}$ will yield a new component $\textbf{w}'_{b0}$ , while the $\textbf{k}_w$ and $\textbf{w}_{b0}$ produce the $\textbf{w}'_{e0}$ . All of these torque components belong to the transverse waves.
Moreover, the amplitudes of torque wave components, $\textbf{w}'_{e0}$ and $\textbf{w}'_{b0}$ , are equal to that of $\textbf{w}_{e0}$ and $\textbf{w}_{b0}$ respectively.

\psubsection{Force wave equation}

In a similar way to the case of torque wave, we can study some wave features about the force in the gravitational field. In the quaternion space, we choose the mass continuity equation as the gauge equation from Eq.(69) to simplify the following calculation, that is $n_0 = 0$.

The force density $\textbf{n}$ can be separated into three components, $\textbf{n} = \textbf{n}_e / v_0 + \textbf{n}_b + \textbf{n}_s$ ,
\begin{eqnarray}
& \textbf{n}_e / v_0 = v_0 \partial_0 \textbf{w} + v_0 \nabla w_0~,~ \textbf{n}_b = v_0 \nabla \times \textbf{w}~,
\\
& \textbf{n}_s = w_0 (\textbf{g} / v_0 + \textbf{b}) + (\textbf{g} / v_0 + \textbf{b}) \times \textbf{w}~.
\end{eqnarray}

\psubsubsection{Case A}

Similar to Eqs.(73) and (74), in case of $n_0 = 0$ and $\textbf{n}_s = 0$, the definition equation of quaternion physical quantity, Eq.(70), can be decomposed into the Maxwell-like equations,
\begin{eqnarray}
& \nabla \cdot (\textbf{n}_b / v_0) = 0~, ~\partial_0 (\textbf{n}_b / v_0) + \nabla^* \times (\textbf{n}_e / v_0) / v_0 = 0 ~,
\\
& \nabla^* \cdot (\textbf{n}_e / v_0) / v_0 = c_0 / v_0~, ~ \partial_0 (\textbf{n}_e / v_0) / v_0 + \nabla^* \times (\textbf{n}_b  / v_0) = \textbf{c} / v_0 ~.
\end{eqnarray}

In the above, the $\textbf{n}_s = 0$ is equivalent to $\textbf{g} / v_0 + \textbf{b} = 0$ from Eq.(93) when $\mathbb{W}_g \neq 0$. If there exists further $\mathbb{B}_g = 0$ and $\mathbb{C}_g = 0$ , the above or Eq.(70) can be reduced to,
\begin{eqnarray}
\lozenge^* \circ \mathbb{N}_g = 0~.
\end{eqnarray}

Applying the operator $\lozenge$ to the above, we get the wave equation
\begin{eqnarray}
(\partial_0^2 + \partial_1^2 + \partial_2^2 + \partial_3^2) \mathbb{N}_g = 0~.
\end{eqnarray}

The above is Laplace equation for the force in the quaternion space too. Proceeding with the operator $\partial_0$ and $\nabla$, we can obtain the wave equation about the force components, $\textbf{n}_e$ and $\textbf{n}_b$, from Eqs.(94) and (95),
\begin{eqnarray}
 (\partial_0^2 + \partial_1^2 + \partial_2^2 + \partial_3^2) \textbf{n}_e = 0~,
~ (\partial_0^2 + \partial_1^2 + \partial_2^2 + \partial_3^2) \textbf{n}_b = 0~.
\end{eqnarray}

The above means that two components of force, $\textbf{n}_e$ and $\textbf{n}_b$, are both possessed of the wave features in the gravitational field. These force component is similar to the periodic force in the vibrational mechanics partially, and may activize the vortex street in the fluid mechanics.

\begin{table}[h]
\caption{Maxwell-like equations for field potential and for torque.}
\label{tab:table2}
\centering
\begin{tabular}{lll}
\hline\hline
& field potential                          &                torque                          \\
\hline
field strength   &  $\textbf{g}/v_0 = \partial_0 \textbf{a} + \nabla a_0 $ & $\textbf{n}_e / v_0 = v_0 \partial_0 \textbf{w} + v_0 \nabla w_0$ \\
$$   &  $ \textbf{b} = \nabla \times \textbf{a}$ &  $\textbf{n}_b = v_0 \nabla \times \textbf{w}$ \\
gauge equation  &  $b_0 = 0$ &  $n_0 = 0$ \\
$$   &  $ \sim $ &  $\textbf{n}_s = 0$ \\
field equation   &  $- \mu \mathbb{S}_g = (\mathbb{B}_g / v_0 + \lozenge)^* \circ \mathbb{B}_g~~$ & $- \mu \mathbb{C}_g = ( \mathbb{B}_g/v_0 + \lozenge)^* \circ \mathbb{N}_g$  \\
$$   &  $\mu = \mu_g $ & $\mu = - 1/ v_0$ \\
Maxwell-like equations   &  $\nabla \cdot \textbf{b} = 0$  &  $\nabla \cdot (\textbf{n}_b / v_0) = 0$ \\
$$   &  $\partial_0 \textbf{b} + \nabla^* \times \textbf{g} / v_0 = 0$  &  $\partial_0 (\textbf{n}_b / v_0) + \nabla^* \times (\textbf{n}_e / v_0) / v_0 = 0$  \\
$$   &  $\nabla^* \cdot \textbf{g} = - \widehat{m} / \varepsilon_g$  & $\nabla^* \cdot (\textbf{n}_e / v_0) / v_0 = c_0 / v_0$  \\
$$   &  $\partial_0 \textbf{g} / v_0 + \nabla^* \times \textbf{b} = - \mu_g \textbf{s}$  & $\partial_0 (\textbf{n}_e / v_0) / v_0 + \nabla^* \times (\textbf{n}_b  / v_0) = \textbf{c} / v_0$ \\
\hline\hline
\end{tabular}
\end{table}

\psubsubsection{Case B}

In case of $n_0 = 0$ and $\textbf{n}_s \neq 0$, Eq.(70) can be decomposed as follows,
\begin{eqnarray}
& c_0 / v_0 = \nabla^* \cdot \textbf{n} + (\textbf{g}/v_0 + \textbf{b})^* \cdot \textbf{n} / v_0 ~,
\\
& \textbf{c} / v_0 = \partial_0 \textbf{n} + \nabla^* \times \textbf{n} + (\textbf{g}/v_0 + \textbf{b})^* \times \textbf{n} / v_0 ~.
\end{eqnarray}

In general the above can be separated into the equations more complex than Eqs.(94) and (95), and may possess more complicated wave features than which in Eqs.(97) and (98).

Specially there may exist following conditions,
\begin{eqnarray}
& \nabla^* \cdot ( \textbf{n}_s + \textbf{n}_b ) + (\textbf{g}/v_0 + \textbf{b})^* \cdot \textbf{n} / v_0 = 0 ~,
\\
& \partial_0 ( \textbf{n}_s + \textbf{n}_b ) + \nabla^* \times ( \textbf{n}_e / v_0 + \textbf{n}_s) + (\textbf{g}/v_0 + \textbf{b})^* \times \textbf{n} / v_0 = 0 ~,
\end{eqnarray}
and considering Eq.(94), the above conditions can be reduced to,
\begin{eqnarray}
& \nabla^* \cdot \textbf{n}_s  + (\textbf{g}/v_0 + \textbf{b})^* \cdot \textbf{n} / v_0 = 0 ~,
\\
& \partial_0 \textbf{n}_s + \nabla^* \times \textbf{n}_s + (\textbf{g}/v_0 + \textbf{b})^* \times \textbf{n} / v_0 = 0 ~.
\end{eqnarray}

Under the above conditions when $\textbf{n}_s \neq 0$, Eqs.(99) and (100) can be simplified to the Maxwell-like equations, which is same as Eqs.(94) and (95).

If there exists additionally $\mathbb{C}_g = 0$ , Eqs.(99) and (100) can further be reduced to,
\begin{eqnarray}
\lozenge^* \circ \mathbb{N}_g = 0~,
\end{eqnarray}
and then we have Eqs.(97) and (98), and some similar inferences about the force components.

In Eq.(70), when there exists the condition relation $\mathbb{C}_g - \mathbb{B}_g^* \circ \mathbb{N}_g = 0$ , we should gain some deductions more complex than the above directly. And the alternating field strength, torque waves, and force waves etc may cause jointly the density waves in the galactic dynamics of astronomy.

\psubsubsection{Transverse wave}

For the force wave with angular frequency $\omega$, the force wave should be a harmonic function $cos \omega t$ as well,  and then should be chosen as the function $exp(-i \omega t)$ . The force wave $\textbf{n}_e$ and $\textbf{n}_b$ can be written as follows,
\begin{eqnarray}
\textbf{n}_e = \textbf{n}_e (r) exp(- i \omega t ) ~,~ \textbf{n}_b = \textbf{n}_b (r) exp(- i \omega t ) ~,
\end{eqnarray}
combining this with Eq.(97), gives
\begin{eqnarray}
\left\{ - ( \omega / v_0 )^2 + \Sigma \partial_j^2 \right\} \textbf{n}_e (r) = 0~, ~ \left\{ - ( \omega / v_0 )^2 + \Sigma \partial_j^2 \right\} \textbf{n}_b (r) = 0~.
\end{eqnarray}

From the analysis of Eqs.(94) and (95), we find that the force wave should be one hyperbolic cosine $cos ( i \gamma)$ as well, and then should be chosen as the function $exp(-i \textbf{i} \gamma)$ . The wave-like vector $\textbf{k}_n = \Sigma (\emph{\textbf{i}}_j k_{nj})$ , while the $ \gamma = - \Sigma ( r_j k_{nj} ) $ , with $k_{nj}$ being the coefficient.

The force wave $\textbf{n}_e (r)$ and $\textbf{n}_b (r)$ are
\begin{eqnarray}
\textbf{n}_e (r) = \textbf{n}_{e0} \circ exp(-i \textbf{i} \gamma) ~,
~ \textbf{n}_b (r) = \textbf{n}_{b0} \circ exp(-i \textbf{i} \gamma) ~,
\end{eqnarray}
where $\textbf{n}_{e0}$ and $\textbf{n}_{b0}$ both are constant vectors in the quaternion space.

And then we have the result from the above and Eq.(106) ,
\begin{eqnarray}
- (\omega / v_0 )^2 + \Sigma k_{nj}^2 = 0~.
\end{eqnarray}

In the gravitational field, we will find that force waves are the transverse waves in a vacuum. Similar to Eqs.(89)$\sim$(91), there are following results from Eqs.(94) and (95),
\begin{eqnarray}
& \textbf{k}_n \cdot \textbf{n}_{e0} = 0~, ~\textbf{k}_n \cdot \textbf{n}'_{e0} = 0~, ~\textbf{k}_n \cdot \textbf{n}_{b0} = 0~, ~\textbf{k}_n \cdot \textbf{n}'_{b0} = 0~,\\
& \textbf{k}_n \times \textbf{n}'_{e0} + \omega \textbf{n}_{b0} = 0~,
~\textbf{k}_n \times \textbf{n}_{e0} - \omega \textbf{n}'_{b0} = 0~,\\
&\textbf{k}_n \times \textbf{n}'_{b0} + \omega (\textbf{n}_{e0} / v_0) / v_0 = 0~,
~\textbf{k}_n \times \textbf{n}_{b0} - \omega (\textbf{n}'_{e0} / v_0) / v_0 = 0~,
\end{eqnarray}
where $\textbf{n}'_{e0} = \textbf{n}_{e0} \circ \textbf{i}$ , $\textbf{n}'_{b0} = \textbf{n}_{b0} \circ \textbf{i}$ .

The above means that there are some relationships between $\mathbb{N}_g$ with $\mathbb{N}_g \circ \textbf{i}$ . The $\textbf{k}_n$ and $\textbf{n}_{e0}$ will yield a new component $\textbf{n}'_{b0}$ , while the $\textbf{k}_n$ and $\textbf{n}_{b0}$ produce the $\textbf{n}'_{e0}$ . All of these force components belong to the transverse waves.
Moreover, the amplitudes of force wave components, $\textbf{n}'_{e0}$ and $\textbf{n}'_{b0}$ , are the same as that of $\textbf{n}_{e0}$ and $\textbf{n}_{b0}$ respectively.

\begin{table}[h]
\caption{The physical quantity and related waves in the gravitational field.}
\label{tab:table2}
\centering
\begin{tabular}{lccc}
\hline\hline
& field strength  &  torque   &    force   \\
\hline
static quantity          &  $yes$  &  $yes$  &  $yes$ \\
alternating quantity     &  $yes$  &  $yes$  &  $yes$ \\
independent wave         &  $yes$  &  $yes$  &  $yes$ \\
\hline\hline
\end{tabular}
\end{table}

\psection{Torque wave and force wave in electromagnetic field}

The torque wave feature as well as the force wave feature can be described simultaneously by the algebra of octonions in the gravitational field and electromagnetic field.

In the case for coexistence of electromagnetic field and gravitational field, the octonion angular momentum density
$\mathbb{L} = \Sigma (l_i \emph{\textbf{i}}_i + L_i \emph{\textbf{I}}_i)$ is defined from the octonion radius vector $\mathbb{R}$ and octonion field source $\mathbb{S}$ in Eq.(24).
\begin{eqnarray}
\mathbb{L} = (\mathbb{R} + k_{rx} \mathbb{X} ) \circ \mathbb{P}~,
\end{eqnarray}
with
\begin{eqnarray}
l_0 = && (r_0 + k_{rx} x_0) p_0 + (\textbf{R}_0 + k_{rx} \textbf{X}_0) \circ \textbf{P}_0 + (\textbf{r} + k_{rx} \textbf{x}) \cdot \textbf{p} + (\textbf{R} + k_{rx} \textbf{X}) \cdot \textbf{P}~,
\nonumber \\
\textbf{l} = && (r_0 + k_{rx} x_0) \textbf{p} + (\textbf{R} + k_{rx} \textbf{X}) \circ \textbf{P}_0
+ (\textbf{r} + k_{rx} \textbf{x}) \times \textbf{p} + (\textbf{R}_0 + k_{rx} \textbf{X}_0) \circ \textbf{P}
\nonumber\\
&& + p_0 (\textbf{r} + k_{rx} \textbf{x}) + (\textbf{R} + k_{rx} \textbf{X}) \times \textbf{P}~,
\nonumber \\
\textbf{L}_0 = && (r_0 + k_{rx} x_0) \textbf{P}_0 + (\textbf{r} + k_{rx} \textbf{x}) \cdot \textbf{P}
+ p_0 (\textbf{R}_0 + k_{rx} \textbf{X}_0) + (\textbf{R} + k_{rx} \textbf{X}) \cdot \textbf{p}~,
\nonumber \\
\textbf{L} = && (r_0 + k_{rx} x_0) \textbf{P} + (\textbf{r} + k_{rx} \textbf{x}) \circ \textbf{P}_0
+ (\textbf{r} + k_{rx} \textbf{x}) \times \textbf{P} + (\textbf{R}_0 + k_{rx} \textbf{X}_0) \circ \textbf{p}
\nonumber\\
&& + p_0 (\textbf{R} + k_{rx} \textbf{X}) + (\textbf{R} + k_{rx} \textbf{X}) \times \textbf{p}~,
\nonumber
\end{eqnarray}
where $\mathbb{P} = \mu \mathbb{S} / \mu_g$. $\textbf{L}_0 = L_0 \emph{\textbf{I}}_0$, $\textbf{L} = \Sigma (L_j \emph{\textbf{I}}_j)$. $\textbf{P}_0 = P_0 \emph{\textbf{I}}_0 $, $\textbf{P} = \Sigma (P_j \emph{\textbf{I}}_j )$. $\textbf{R}_0 = R_0 \emph{\textbf{I}}_0 $, $\textbf{R} = \Sigma (R_j \emph{\textbf{I}}_j )$. $\mathbb{X} = \Sigma (x_i \emph{\textbf{i}}_i + X_i \emph{\textbf{I}}_i)$. $\textbf{X}_0 = X_0 \emph{\textbf{I}}_0$, $\textbf{X} = \Sigma (X_j \emph{\textbf{I}}_j )$. The derivation of octonion physical quantity $\mathbb{X}$ yields the gravitational potential as well as the electromagnetic potential.

The octonion energy-torque density $\mathbb{W} = \Sigma (w_i \emph{\textbf{i}}_i + W_i \emph{\textbf{I}}_i )$ can be defined from the octonion field strength $\mathbb{B}$ and octonion angular momentum density $\mathbb{L}$ in Eq.(113).
\begin{eqnarray}
\mathbb{W} = v_0 ( \mathbb{B}/v_0 + \lozenge) \circ \mathbb{L}~,
\end{eqnarray}
with
\begin{eqnarray}
w_0 = && v_0 \partial_0 l_0 + v_0 \nabla \cdot \textbf{l} + (\textbf{g} / v_0 + \textbf{b}) \cdot \textbf{l}
+ k_{eg} (\textbf{E} / V_0 + \textbf{B}) \cdot \textbf{L}~,
\nonumber
\\
\textbf{w} = && v_0 \partial_0 \textbf{l} + v_0 \nabla l_0 + v_0 \nabla \times \textbf{l} + l_0 (\textbf{g} / v_0 + \textbf{b}) + (\textbf{g} / v_0 + \textbf{b}) \times \textbf{l}
\nonumber\\
&& + k_{eg} (\textbf{E} / V_0 + \textbf{B}) \times \textbf{L}
+ k_{eg} (\textbf{E} / V_0 + \textbf{B}) \circ \textbf{L}_0 ~,
\nonumber\\
\textbf{W}_0 = && v_0 \partial_0 \textbf{L}_0 + k_{eg} (\textbf{E} / V_0 + \textbf{B}) \cdot \textbf{l}
+ v_0 \nabla \cdot \textbf{L} + (\textbf{g} / v_0 + \textbf{b}) \cdot \textbf{L}~,
\nonumber\\
\textbf{W} = && v_0 \partial_0 \textbf{L} + l_0 k_{eg} (\textbf{E} / V_0 + \textbf{B}) + v_0 \nabla \times \textbf{L}
+ v_0 \nabla \circ \textbf{L}_0 + k_{eg} (\textbf{E} / V_0 + \textbf{B}) \times \textbf{l}
\nonumber\\
&& + (\textbf{g} / v_0 + \textbf{b}) \times \textbf{L}
+ (\textbf{g} / v_0 + \textbf{b}) \circ \textbf{L}_0~,
\nonumber
\end{eqnarray}
where $-w_0/2$ is the energy density, includes the density of kinetic energy, potential energy, field energy, work, electric potential energy, magnetic potential energy, and interacting energy between dipole moment with electromagnetic strength etc. $\textbf{w}/2 = \Sigma (w_j \emph{\textbf{i}}_j )/2$ is the torque density, includes that caused by gravitational force, electromagnetic force, and other force terms etc. $\textbf{W}_0 = W_0 \emph{\textbf{I}}_0$, $\textbf{W} = \Sigma (W_j \emph{\textbf{I}}_j)$. $\textbf{A}_0 + \textbf{A} = k_{eg} \mathbb{A}_e$. $\textbf{A}_0 = V_0 (\partial_0 \textbf{X}_0 + \nabla \cdot \textbf{X}$) and $\textbf{A} = V_0 (\partial_0 \textbf{X} + \nabla \circ \textbf{X}_0 + \nabla \times \textbf{X})$ are related to the scalar potential and vectorial potential of electromagnetic field respectively.

The octonion power-force density $\mathbb{N} = \Sigma (n_i \emph{\textbf{i}}_i + N_i \emph{\textbf{I}}_i)$ is defined from the octonion energy-torque density $\mathbb{W}$ in Eq.(114) and field strength $\mathbb{B}$ ,
\begin{eqnarray}
\mathbb{N} = v_0 ( \mathbb{B}/v_0 + \lozenge)^* \circ \mathbb{W}~,
\end{eqnarray}
with
\begin{eqnarray}
n_0 = && v_0 \partial_0 w_0 + v_0 \nabla^*  \cdot \textbf{w} + (\textbf{g} / v_0 + \textbf{b})^*  \cdot \textbf{w}
+ k_{eg} (\textbf{E} / V_0 + \textbf{B})^*  \cdot \textbf{W}~,
\nonumber
\\
\textbf{n} = && v_0 \partial_0 \textbf{w} + v_0 \nabla^* w_0 + v_0 \nabla^*  \times \textbf{w} + w_0 (\textbf{g} / v_0 + \textbf{b})^*  + (\textbf{g} / v_0 + \textbf{b})^*  \times \textbf{w}
\nonumber\\
&& + k_{eg} (\textbf{E} / V_0 + \textbf{B})^*  \times \textbf{W}
+ k_{eg} (\textbf{E} / V_0 + \textbf{B})^*  \circ \textbf{W}_0 ~,
\nonumber\\
\textbf{N}_0 = && v_0 \partial_0 \textbf{W}_0 + k_{eg} (\textbf{E} / V_0 + \textbf{B})^*  \cdot \textbf{w}
+ v_0 \nabla^*  \cdot \textbf{W} + (\textbf{g} / v_0 + \textbf{b})^*  \cdot \textbf{W}~,
\nonumber\\
\textbf{N} = && v_0 \partial_0 \textbf{W} + w_0 k_{eg} (\textbf{E} / V_0 + \textbf{B})^*  + v_0 \nabla^*  \times \textbf{W} + v_0 \nabla^*  \circ \textbf{W}_0
\nonumber\\
&& + k_{eg} (\textbf{E} / V_0 + \textbf{B})^*  \times \textbf{w} + (\textbf{g} / v_0 + \textbf{b})^*  \times \textbf{W} + (\textbf{g} / v_0 + \textbf{b})^*  \circ \textbf{W}_0~,
\nonumber
\end{eqnarray}
where $\textbf{N}_0 = N_0 \emph{\textbf{I}}_0$, $\textbf{N} = \Sigma (N_j \emph{\textbf{I}}_j)$. The $f_0 = - n_0/(2 v_0)$ is the power density, while the vectorial part $\textbf{n} = \Sigma (n_j  \emph{\textbf{i}}_j )$ is the function of forces. The $\textbf{f} = - \textbf{n} / (2 v_0)$ is the force density, which includes the density of inertial force, gravity, Lorentz force, gradient of energy, and interacting force between dipole moment with magnetic strength etc.

The octonion physical quantity $\mathbb{C} = \Sigma (c_i \emph{\textbf{i}}_i + C_i \emph{\textbf{I}}_i)$ is defined from the octonion power-force density $\mathbb{N}$ in Eq.(115) and field strength $\mathbb{B}$ ,
\begin{eqnarray}
\mathbb{C} = v_0 ( \mathbb{B}/v_0 + \lozenge)^* \circ \mathbb{N}~,
\end{eqnarray}
with
\begin{eqnarray}
c_0 = && v_0 \partial_0 n_0 + v_0 \nabla^*  \cdot \textbf{n} + (\textbf{g} / v_0 + \textbf{b})^*  \cdot \textbf{n}
+ k_{eg} (\textbf{E} / V_0 + \textbf{B})^*  \cdot \textbf{N}~,
\nonumber
\\
\textbf{c} = && v_0 \partial_0 \textbf{n} + v_0 \nabla^* n_0 + v_0 \nabla^*  \times \textbf{n} + n_0 (\textbf{g} / v_0 + \textbf{b})^*  + (\textbf{g} / v_0 + \textbf{b})^*  \times \textbf{n}
\nonumber\\
&& + k_{eg} (\textbf{E} / V_0 + \textbf{B})^*  \times \textbf{N}
+ k_{eg} (\textbf{E} / V_0 + \textbf{B})^*  \circ \textbf{N}_0 ~,
\nonumber\\
\textbf{C}_0 = && v_0 \partial_0 \textbf{N}_0 + k_{eg} (\textbf{E} / V_0 + \textbf{B})^*  \cdot \textbf{n}
+ v_0 \nabla^*  \cdot \textbf{N} + (\textbf{g} / v_0 + \textbf{b})^*  \cdot \textbf{N}~,
\nonumber\\
\textbf{C} = && v_0 \partial_0 \textbf{N} + n_0 k_{eg} (\textbf{E} / V_0 + \textbf{B})^*  + v_0 \nabla^*  \times \textbf{N} + v_0 \nabla^*  \circ \textbf{N}_0
\nonumber\\
&& + k_{eg} (\textbf{E} / V_0 + \textbf{B})^*  \times \textbf{n} + (\textbf{g} / v_0 + \textbf{b})^*  \times \textbf{N} + (\textbf{g} / v_0 + \textbf{b})^*  \circ \textbf{N}_0~,
\nonumber
\end{eqnarray}
where $\textbf{C}_0 = C_0 \emph{\textbf{I}}_0$, $\textbf{C} = \Sigma (C_j \emph{\textbf{I}}_j)$.

\psubsection{Torque wave equation}

Comparing with the classical electromagnetic theory, there are the $\mathbb{W}_g$ and $\mathbb{N}_g$ in the $\mathbb{E}_g$ space, while the $\mathbb{W}_e$ and $\mathbb{N}_e$ in the $\mathbb{E}_e$ space. And the physical quantity $\mathbb{W}$ as well as $\mathbb{N}$ are devised to describe consistently in the electromagnetic and gravitational fields.

According to the basis vectors, Eq.(114) can be decomposed further as follows,
\begin{eqnarray}
& (\mathbb{B}_g^* \circ \mathbb{W}_g + k_{eg} \mathbb{B}_e^* \circ \mathbb{W}_e ) / v_0 + \lozenge^* \circ \mathbb{W}_g = \mathbb{N}_g / v_0~,
\\
& (\mathbb{B}_g^* \circ \mathbb{W}_e + k_{eg} \mathbb{B}_e^* \circ \mathbb{W}_g ) / v_0 + \lozenge^* \circ \mathbb{W}_e = \mathbb{N}_e / v_0~,
\end{eqnarray}
where $\mathbb{W}_e = \Sigma (W_i \emph{\textbf{I}}_i)$ , $\mathbb{N}_e = \Sigma (N_i \emph{\textbf{I}}_i)$ . In the above, Eq.(117) is similar to Eq.(69) and is suitable for the torque wave equation, while Eq.(118) is for the torque-like wave equation.

Choosing the gauge equation, $w_0 = 0$, from Eq.(114) to simplify following calculation. The torque density $\textbf{w}$ can be separated into three components, $\textbf{w} = \textbf{w}_e / v_0 + \textbf{w}_b + \textbf{w}_s$ ,
\begin{eqnarray}
& \textbf{w}_e / v_0 = v_0 \partial_0 \textbf{l} + v_0 \nabla l_0~,~ \textbf{w}_b = v_0 \nabla \times \textbf{l}~,
\\
& \textbf{w}_s = l_0 (\textbf{g} / v_0 + \textbf{b}) + (\textbf{g} / v_0 + \textbf{b}) \times \textbf{l}
+ k_{eg} (\textbf{E} / V_0 + \textbf{B}) \times \textbf{L} + k_{eg} (\textbf{E} / V_0 + \textbf{B}) \circ \textbf{L}_0 ~.
\end{eqnarray}

Substituting Eqs.(117), (119), and (120) for Eqs.(69), (71), and (72) respectively, there are some deductions about the torque wave features similar to Eqs.(73)$\sim$(91) in the case for coexistence of gravitational and electromagnetic fields.

\psubsection{Torque-like wave equation}

From Eq.(114) in the octonion space, we choose the gauge equation, $\textbf{W}_0 = 0$. And the torque-like quantity $\textbf{W}$ can be separated into three components, $\textbf{W} = \textbf{W}_e / V_0 + \textbf{W}_b + \textbf{W}_s$ ,
\begin{eqnarray}
& \textbf{W}_e / V_0 = v_0 \partial_0 \textbf{L} + v_0 \nabla \circ \textbf{L}_0~,~ \textbf{W}_b = v_0 \nabla \times \textbf{L}~,
\\
& \textbf{W}_s = l_0 k_{eg} (\textbf{E} / V_0 + \textbf{B}) + (\textbf{g} / v_0 + \textbf{b}) \circ \textbf{L}_0
+ k_{eg} (\textbf{E} / V_0 + \textbf{B}) \times \textbf{l} + (\textbf{g} / v_0 + \textbf{b}) \times \textbf{L}~.
\end{eqnarray}

\psubsubsection{Case A}

Similar to Eqs.(73) and (74), in case of $\textbf{W}_0 = 0$ and $\textbf{W}_s = 0$, Eq.(118) can be decomposed into the Maxwell-like equations,
\begin{eqnarray}
& \nabla \cdot (\textbf{W}_b / v_0) = 0~, ~\partial_0 (\textbf{W}_b / v_0) + \nabla^* \times (\textbf{W}_e / V_0) / v_0 = 0 ~,
\\
& \nabla^* \cdot (\textbf{W}_e / V_0) / v_0 = \textbf{N}_0 / v_0~, ~ \partial_0 (\textbf{W}_e / V_0) / v_0 + \nabla^* \times (\textbf{W}_b / v_0) = \textbf{N} / v_0 ~.
\end{eqnarray}

In the above, the $\textbf{W}_s = 0$ means that $\textbf{g} / v_0 + \textbf{b} = 0$ and $\textbf{E} / V_0 + \textbf{B} = 0$ from Eq.(122) when $\mathbb{L} \neq 0$. If there exists $\mathbb{N}_e = 0$ , the above or Eq.(118) can further be reduced to,
\begin{eqnarray}
\lozenge^* \circ \mathbb{W}_e = 0~.
\end{eqnarray}

Applying the operator $\lozenge$ to the above, we have the wave equation or Laplace equation
\begin{eqnarray}
\lozenge \circ (\lozenge^* \circ \mathbb{W}_e) = 0~, ~or ~ (\partial_0^2 + \partial_1^2 + \partial_2^2 + \partial_3^2) \mathbb{W}_e = 0~.
\end{eqnarray}

Proceeding with the operator $\partial_0$ and $\nabla$, we can obtain the wave equation about the components of torque-like quantity from Eqs.(123) and (124),
\begin{eqnarray}
 (\partial_0^2 + \partial_1^2 + \partial_2^2 + \partial_3^2) \textbf{W}_e = 0~,
~ (\partial_0^2 + \partial_1^2 + \partial_2^2 + \partial_3^2) \textbf{W}_b = 0~.
\end{eqnarray}

The above means that two components of torque-like quantity, $\textbf{W}_e$ and $\textbf{W}_b$, are both possessed of wave features in the electromagnetic field.

\begin{table}[h]
\caption{Maxwell-like equations for torque and torque-like in the electromagnetic field.}
\label{tab:table2}
\centering
\begin{tabular}{lll}
\hline\hline
$$ & torque                          &          torque-like                        \\
\hline
field strength   &  $\textbf{w}_e / v_0 = v_0 \partial_0 \textbf{l} + v_0 \nabla l_0$ & $\textbf{W}_e / V_0 = v_0 \partial_0 \textbf{L} + v_0 \nabla \circ \textbf{L}_0$ \\
$$   &  $\textbf{w}_b = v_0 \nabla \times \textbf{l}$ &  $\textbf{W}_b = v_0 \nabla \times \textbf{L}$ \\
gauge equation  &  $w_0 = 0$ &  $\textbf{W}_0 = 0$ \\
$$   &  $\textbf{w}_s = 0$ &  $\textbf{W}_s = 0$ \\
field equation   &  $\mathbb{N}_g = \mathbb{B}_g^* \circ \mathbb{W}_g + k_{eg} \mathbb{B}_e^* \circ \mathbb{W}_e + v_0 \lozenge^* \circ \mathbb{W}_g$ & $\mathbb{N}_e = \mathbb{B}_g^* \circ \mathbb{W}_e + k_{eg} \mathbb{B}_e^* \circ \mathbb{W}_g + v_0 \lozenge^* \circ \mathbb{W}_e$  \\
$$   &  $\mu = - 1/ v_0$ & $\mu = - 1/ v_0$ \\
Maxwell-like  &  $\nabla \cdot (\textbf{w}_b / v_0) = 0$  &  $\nabla \cdot (\textbf{W}_b / v_0) = 0$ \\
equations   &  $\partial_0 (\textbf{w}_b / v_0) + \nabla^* \times (\textbf{w}_e / v_0) / v_0 = 0$  &  $\partial_0 (\textbf{W}_b / v_0) + \nabla^* \times (\textbf{W}_e / V_0) / v_0 = 0$  \\
$$   &  $\nabla^* \cdot (\textbf{w}_e / v_0) / v_0 = n_0 / v_0$  & $\nabla^* \cdot (\textbf{W}_e / V_0) / v_0 = \textbf{N}_0 / v_0$  \\
$$   &  $\partial_0 (\textbf{w}_e / v_0) / v_0 + \nabla^* \times (\textbf{w}_b  / v_0) = \textbf{n} / v_0$  & $\partial_0 (\textbf{W}_e / V_0) / v_0 + \nabla^* \times (\textbf{W}_b  / v_0) = \textbf{N} / v_0$ \\
\hline\hline
\end{tabular}
\end{table}

\psubsubsection{Case B}

In case of $\textbf{W}_0 = 0$ and $\textbf{W}_s \neq 0$, Eq.(118) can be decomposed as follows,
\begin{eqnarray}
\textbf{N}_0  / v_0 = && \nabla^*  \cdot \textbf{W} + (\textbf{g} / v_0 + \textbf{b})^*  \cdot \textbf{W} / v_0  + k_{eg} (\textbf{E} / V_0 + \textbf{B})^*  \cdot \textbf{w} / v_0 ~,
\\
\textbf{N}  / v_0 = && \partial_0 \textbf{W} + \nabla^*  \times \textbf{W}
+ (\textbf{g} / v_0 + \textbf{b})^*  \times \textbf{W}  / v_0
\nonumber\\
&& + k_{eg} (\textbf{E} / V_0 + \textbf{B})^*  \times \textbf{w} / v_0  + w_0 k_{eg} (\textbf{E} / V_0 + \textbf{B})^* / v_0 ~.
\end{eqnarray}

In general the above can be separated into the equations more complex than Eqs.(123) and (124), and may possess more complicated wave features than which in Eqs.(126) and (127).

Specially there may exist following conditions,
\begin{eqnarray}
0 = && \nabla^* \cdot ( \textbf{W}_s + \textbf{W}_b ) + (\textbf{g}/v_0 + \textbf{b})^* \cdot \textbf{W} / v_0 + k_{eg} (\textbf{E} / V_0 + \textbf{B})^*  \cdot \textbf{w} / v_0 ~,
\\
0 = && \partial_0 ( \textbf{W}_s + \textbf{W}_b ) + \nabla^* \times ( \textbf{W}_e / V_0 + \textbf{W}_s) + (\textbf{g}/v_0 + \textbf{b})^* \times \textbf{W} / v_0
\nonumber\\
&& + k_{eg} (\textbf{E} / V_0 + \textbf{B})^*  \times \textbf{w} / v_0  + w_0 k_{eg} (\textbf{E} / V_0 + \textbf{B})^*  / v_0 ~,
\end{eqnarray}
and considering Eq.(123), the above conditions can be reduced to,
\begin{eqnarray}
0 = && \nabla^* \cdot \textbf{W}_s + (\textbf{g}/v_0 + \textbf{b})^* \cdot \textbf{W} / v_0 + k_{eg} (\textbf{E} / V_0 + \textbf{B})^*  \cdot \textbf{w} / v_0 ~,
\\
0 = && \partial_0 \textbf{W}_s + \nabla^* \times \textbf{W}_s + (\textbf{g}/v_0 + \textbf{b})^* \times \textbf{W} / v_0 + k_{eg} (\textbf{E} / V_0 + \textbf{B})^*  \times \textbf{w} / v_0
\nonumber\\
&& + w_0 k_{eg} (\textbf{E} / V_0 + \textbf{B})^*  / v_0 ~,
\end{eqnarray}

Under the above conditions, from Eqs.(128) and (129) when $\textbf{W}_s \neq 0$, we can obtain the Maxwell-like equations, which are same as Eqs.(123) and (124).

If there exists additionally $\mathbb{N}_e = 0$ , Eqs.(128) and (129) can further be reduced to,
\begin{eqnarray}
\lozenge^* \circ \mathbb{W}_e = 0~,
\end{eqnarray}
and then we have Eqs.(126) and (127), and some similar inferences about torque-like components.

In Eq.(118), when there exists the condition relation $\mathbb{N}_e - (\mathbb{B}_g^* \circ \mathbb{W}_e + k_{eg} \mathbb{B}_e^* \circ \mathbb{W}_g ) = 0$ , we should gain some more complex deductions than the above directly.

\psubsubsection{Transverse wave}

For the torque wave with angular frequency $\omega$, the torque wave should be a harmonic function $cos \omega t$ ,  and then should be chosen as the function $exp(-i \omega t)$ . The torque wave $\textbf{W}_e$ and $\textbf{W}_b$ can be written as follows,
\begin{eqnarray}
\textbf{W}_e = \textbf{W}_e (r) exp(- i \omega t ) ~,~ \textbf{W}_b = \textbf{W}_b (r) exp(- i \omega t ) ~,
\end{eqnarray}
combining this with Eq.(126), gives
\begin{eqnarray}
\left\{ - ( \omega / v_0 )^2 + \Sigma \partial_j^2 \right\} \textbf{W}_e (r) = 0~, ~ \left\{ - ( \omega / v_0 )^2 + \Sigma \partial_j^2 \right\} \textbf{W}_b (r) = 0~.
\end{eqnarray}

From the analysis of Eqs.(123) and (124), we find that the torque wave should be one hyperbolic cosine $cos ( i \beta)$ , and be chosen as the function $exp(-i \textbf{I} \beta)$ . The wave-like vector $\textbf{K}_w = \Sigma (\emph{\textbf{I}}_j K_{wj})$ , while the $ \beta = - \Sigma ( r_j K_{wj} ) $ , with $K_{wj}$ being the coefficient.

The torque wave $\textbf{W}_e (r)$ and $\textbf{W}_b (r)$ are
\begin{eqnarray}
\textbf{W}_e (r) = \textbf{W}_{e0} \circ exp(-i \textbf{I} \beta) ~,
~ \textbf{W}_b (r) = \textbf{W}_{b0} \circ exp(-i \textbf{I} \beta) ~,
\end{eqnarray}
where $\textbf{W}_{e0}$ and $\textbf{W}_{b0}$ both are constant vectors in the octonion space.

And then we have the result from the above and Eq.(135) ,
\begin{eqnarray}
- (\omega / v_0 )^2 + \Sigma K_{wj}^2 = 0~.
\end{eqnarray}

In the electromagnetic field, we will find that torque waves are the transverse waves in a vacuum. Similar to Eqs.(110)$\sim$(112), there are following results from Eqs.(123) and (124),
\begin{eqnarray}
& \textbf{K}_w \cdot \textbf{W}_{e0} = 0~, ~\textbf{K}_w \cdot \textbf{W}'_{e0} = 0~, ~\textbf{K}_w \cdot \textbf{W}_{b0} = 0~, ~\textbf{K}_w \cdot \textbf{W}'_{b0} = 0~,\\
& \textbf{K}_w \times \textbf{W}'_{e0} + k_v \omega \textbf{W}_{b0} = 0~,
~\textbf{K}_w \times \textbf{W}_{e0} - k_v \omega \textbf{W}'_{b0} = 0~,\\
& \textbf{K}_w \times \textbf{W}'_{b0} + \omega (\textbf{W}_{e0} / V_0 ) / v_0 = 0~,
~ \textbf{K}_w \times \textbf{W}_{b0} - \omega (\textbf{W}'_{e0} / V_0 ) / v_0 = 0~,
\end{eqnarray}
where $\textbf{W}'_{e0} = \textbf{W}_{e0} \circ \textbf{I}$ , $\textbf{W}'_{b0} = \textbf{W}_{b0} \circ \textbf{I}$ .

The above means that there are some relationships between $\mathbb{W}_e$ with $\mathbb{W}_e \circ \textbf{I}$ . The $\textbf{K}_w$ and $\textbf{W}_{e0}$ will yield a new component $\textbf{W}'_{b0}$ , while the $\textbf{K}_w$ and $\textbf{W}_{b0}$ produce the $\textbf{W}'_{e0}$ . All of these torque components belong to the transverse waves.
Moreover, the amplitudes of torque wave components, $\textbf{W}'_{e0}$ and $\textbf{W}'_{b0}$ , are the same as that of $\textbf{W}_{e0}$ and $\textbf{W}_{b0}$ respectively.

\psubsection{Force wave equation}

Similar to Eq.(115), the $\mathbb{C}$ is devised to describe the octonion physical quantity consistently in the electromagnetic field and gravitational field. According to the basis vectors, Eq.(116) can be decomposed further as follows,
\begin{eqnarray}
& (\mathbb{B}_g^* \circ \mathbb{N}_g + k_{eg} \mathbb{B}_e^* \circ \mathbb{N}_e ) / v_0 + \lozenge^* \circ \mathbb{N}_g = \mathbb{C}_g / v_0~,
\\
& (\mathbb{B}_g^* \circ \mathbb{N}_e + k_{eg} \mathbb{B}_e^* \circ \mathbb{N}_g ) / v_0 + \lozenge^* \circ \mathbb{N}_e = \mathbb{C}_e / v_0~,
\end{eqnarray}
where $\mathbb{C}_e = \Sigma (C_i \emph{\textbf{I}}_i)$ . In the above, Eq.(142) is similar to Eq.(117) and is suitable for the force wave equation, while Eq.(143) is for the force-like wave equation.

To simplify the following calculation, we choose the mass continuity equation from Eq.(115) as the gauge equation in the electromagnetic field, that is $n_0 = 0$ . The force density $\textbf{n}$ can be divided into three components, $\textbf{n} = \textbf{n}_e / v_0 + \textbf{n}_b + \textbf{n}_s$ ,
\begin{eqnarray}
& \textbf{n}_e / v_0 = v_0 \partial_0 \textbf{w} + v_0 \nabla^* w_0,~ \textbf{n}_b = v_0 \nabla^* \times \textbf{w},
\\
& \textbf{n}_s = w_0 (\textbf{g} / v_0 + \textbf{b})^* + (\textbf{g} / v_0 + \textbf{b})^* \times \textbf{w}
+ k_{eg} (\textbf{E} / V_0 + \textbf{B})^* \times \textbf{W} + k_{eg} (\textbf{E} / V_0 + \textbf{B})^* \circ \textbf{W}_0.
\end{eqnarray}

Substituting Eqs.(142), (144), and (145) for Eqs.(70), (92), and (93) respectively, there are some deductions about the force wave features similar to Eqs.(94)$\sim$(112) in the case for coexistence of gravitational and electromagnetic fields.

\psubsection{Force-like wave equation}

From Eq.(115) in the case for coexistence of electromagnetic and gravitational fields, we choose the charge continuity equation as the gauge equation, that is $\textbf{N}_0 = 0$. And the force-like quantity $\textbf{N}$ can be separated into three components, $\textbf{N} = \textbf{N}_e / V_0 + \textbf{N}_b + \textbf{N}_s$ ,
\begin{eqnarray}
& \textbf{N}_e / V_0 = v_0 \partial_0 \textbf{W} + v_0 \nabla^* \circ \textbf{W}_0,~ \textbf{N}_b = v_0 \nabla^* \times \textbf{W},
\\
& \textbf{N}_s = w_0 k_{eg} (\textbf{E} / V_0 + \textbf{B})^* + (\textbf{g} / v_0 + \textbf{b})^* \circ \textbf{W}_0
+ k_{eg} (\textbf{E} / V_0 + \textbf{B})^* \times \textbf{w} + (\textbf{g} / v_0 + \textbf{b})^* \times \textbf{W}.
\end{eqnarray}

\begin{table}[h]
\caption{Maxwell-like equations for force and force-like in the electromagnetic field.}
\label{tab:table2}
\centering
\begin{tabular}{lll}
\hline\hline
$$ & force                          &          force-like                        \\
\hline
field strength   &  $\textbf{n}_e / v_0 = v_0 \partial_0 \textbf{w} + v_0 \nabla w_0$ & $\textbf{N}_e / V_0 = v_0 \partial_0 \textbf{W} + v_0 \nabla \circ \textbf{W}_0$ \\
$$   &  $\textbf{n}_b = v_0 \nabla \times \textbf{w}$ &  $\textbf{N}_b = v_0 \nabla \times \textbf{W}$ \\
gauge equation  &  $n_0 = 0$ &  $\textbf{N}_0 = 0$ \\
$$   &  $\textbf{n}_s = 0$ &  $\textbf{N}_s = 0$ \\
field equation   &  $\mathbb{C}_g = \mathbb{B}_g^* \circ \mathbb{N}_g + k_{eg} \mathbb{B}_e^* \circ \mathbb{N}_e + v_0 \lozenge^* \circ \mathbb{N}_g$ & $\mathbb{C}_e = \mathbb{B}_g^* \circ \mathbb{N}_e + k_{eg} \mathbb{B}_e^* \circ \mathbb{N}_g + v_0 \lozenge^* \circ \mathbb{N}_e$  \\
$$   &  $\mu = - 1/ v_0$ & $\mu = - 1/ v_0$ \\
Maxwell-like    &  $\nabla \cdot (\textbf{n}_b / v_0) = 0$  &  $\nabla \cdot (\textbf{N}_b / v_0) = 0$ \\
equations   &  $\partial_0 (\textbf{n}_b / v_0) + \nabla^* \times (\textbf{n}_e / v_0) / v_0 = 0$  &  $\partial_0 (\textbf{N}_b / v_0) + \nabla^* \times (\textbf{N}_e / V_0) / v_0 = 0$  \\
$$   &  $\nabla^* \cdot (\textbf{n}_e / v_0) / v_0 = c_0 / v_0$  & $\nabla^* \cdot (\textbf{N}_e / V_0) / v_0 = \textbf{C}_0 / v_0$  \\
$$   &  $\partial_0 (\textbf{n}_e / v_0) / v_0 + \nabla^* \times (\textbf{n}_b  / v_0) = \textbf{c} / v_0$  & $\partial_0 (\textbf{N}_e / V_0) / v_0 + \nabla^* \times (\textbf{N}_b  / v_0) = \textbf{C} / v_0$ \\
\hline\hline
\end{tabular}
\end{table}

\psubsubsection{Case A}

Similar to Eqs.(94) and (95), in case of $\textbf{N}_0 = 0$ and $\textbf{N}_s = 0$, Eq.(143) can be decomposed into the Maxwell-like equations,
\begin{eqnarray}
& \nabla \cdot (\textbf{N}_b / v_0) = 0~, ~\partial_0 (\textbf{N}_b / v_0) + \nabla^* \times (\textbf{N}_e / V_0) / v_0 = 0 ~,
\\
& \nabla^* \cdot (\textbf{N}_e / V_0) / v_0 = \textbf{C}_0 / v_0~, ~ \partial_0 (\textbf{N}_e / V_0) / v_0 + \nabla^* \times (\textbf{N}_b / v_0) = \textbf{C} / v_0 ~.
\end{eqnarray}

In the above, the $\textbf{N}_s = 0$ means that $\textbf{g} / v_0 + \textbf{b} = 0$ and $\textbf{E} / V_0 + \textbf{B} = 0$ from Eq.(147) when $\mathbb{W} \neq 0$. If there exists $\mathbb{C}_e = 0$ , the above or Eq.(143) can further be reduced to,
\begin{eqnarray}
\lozenge^* \circ \mathbb{N}_e = 0~.
\end{eqnarray}

Applying the operator $\lozenge$ to the above, we have the wave equation or Laplace equation
\begin{eqnarray}
\lozenge \circ (\lozenge^* \circ \mathbb{N}_e) = 0~, ~or ~ (\partial_0^2 + \partial_1^2 + \partial_2^2 + \partial_3^2) \mathbb{N}_e = 0~.
\end{eqnarray}

Proceeding with the operator $\partial_0$ and $\nabla$, we can obtain the wave equation about the components of force-like quantity from Eqs.(148) and (149),
\begin{eqnarray}
 (\partial_0^2 + \partial_1^2 + \partial_2^2 + \partial_3^2) \textbf{N}_e = 0~,
~ (\partial_0^2 + \partial_1^2 + \partial_2^2 + \partial_3^2) \textbf{N}_b = 0~.
\end{eqnarray}

The above means that two components of force-like quantity, $\textbf{N}_e$ and $\textbf{N}_b$, are both possessed of wave features in the electromagnetic field.

\psubsubsection{Case B}

In case of $\textbf{N}_0 = 0$ and $\textbf{N}_s \neq 0$, Eq.(143) can be decomposed as follows,
\begin{eqnarray}
\textbf{C}_0 / v_0 = && \nabla^*  \cdot \textbf{N} + (\textbf{g} / v_0 + \textbf{b})^*  \cdot \textbf{N} / v_0  + k_{eg} (\textbf{E} / V_0 + \textbf{B})^*  \cdot \textbf{n} / v_0 ~,
\\
\textbf{C} / v_0 = && \partial_0 \textbf{N} + \nabla^*  \times \textbf{N}
+ (\textbf{g} / v_0 + \textbf{b})^*  \times \textbf{N}  / v_0
\nonumber\\
&& + k_{eg} (\textbf{E} / V_0 + \textbf{B})^*  \times \textbf{n} / v_0  + n_0 k_{eg} (\textbf{E} / V_0 + \textbf{B})^* / v_0 ~.
\end{eqnarray}

In general the above can be separated into more complex equations than Eqs.(148) and (149), and may possess more complicated wave features than that in Eqs.(151) and (152).

Specially there may exist following conditions,
\begin{eqnarray}
0 = && \nabla^* \cdot ( \textbf{N}_s + \textbf{N}_b ) + (\textbf{g}/v_0 + \textbf{b})^* \cdot \textbf{N} / v_0 + k_{eg} (\textbf{E} / V_0 + \textbf{B})^*  \cdot \textbf{n} / v_0 ~,
\\
0 = && \partial_0 ( \textbf{N}_s + \textbf{N}_b ) + \nabla^* \times ( \textbf{N}_e / V_0 + \textbf{N}_s) + (\textbf{g}/v_0 + \textbf{b})^* \times \textbf{N} / v_0
\nonumber\\
&& + k_{eg} (\textbf{E} / V_0 + \textbf{B})^*  \times \textbf{n} / v_0  + n_0 k_{eg} (\textbf{E} / V_0 + \textbf{B})^*  / v_0 ~,
\end{eqnarray}
and considering Eq.(148), the above conditions can be reduced to,
\begin{eqnarray}
0 = && \nabla^* \cdot \textbf{N}_s + (\textbf{g}/v_0 + \textbf{b})^* \cdot \textbf{N} / v_0 + k_{eg} (\textbf{E} / V_0 + \textbf{B})^*  \cdot \textbf{n} / v_0 ~,
\\
0 = && \partial_0 \textbf{N}_s + \nabla^* \times \textbf{N}_s + (\textbf{g}/v_0 + \textbf{b})^* \times \textbf{N} / v_0 + k_{eg} (\textbf{E} / V_0 + \textbf{B})^*  \times \textbf{n} / v_0
\nonumber\\
&& + n_0 k_{eg} (\textbf{E} / V_0 + \textbf{B})^*  / v_0 ~,
\end{eqnarray}

Under the above conditions, from Eqs.(153) and (154) when $\textbf{N}_s \neq 0$, we can obtain the Maxwell-like equations, which are same as Eqs.(148) and (149).

If there exists additionally $\mathbb{C}_e = 0$ , Eqs.(153) and (154) can further be reduced to,
\begin{eqnarray}
\lozenge^* \circ \mathbb{N}_e = 0~,
\end{eqnarray}
and then we have Eqs.(151) and (152), and some similar inferences about force-like components.

In Eq.(143), when there exists the condition relation $\mathbb{C}_e - (\mathbb{B}_g^* \circ \mathbb{N}_e + k_{eg} \mathbb{B}_e^* \circ \mathbb{N}_g ) = 0$ , we should gain some more complex deductions than the above directly.

In the case for coexistence of gravitational and electromagnetic fields, the alternating field strength, torque waves, torque-like waves, force waves, and force-like waves etc may agitate jointly the density waves in the galactic dynamics of astronomy.

\begin{table}[h]
\caption{The physical quantity and their waves in the electromagnetic field.}
\label{tab:table2}
\centering
\begin{tabular}{lccccc}
\hline\hline
& field strength  &  torque     &  torque-like   &    force    &    force-like   \\
\hline
static quantity          &  $yes$  &  $yes$  &  $yes$   &  $yes$  &  $yes$   \\
alternating quantity     &  $yes$  &  $yes$  &  $yes$   &  $yes$  &  $yes$   \\
independent wave         &  $yes$  &  $yes$  &  $yes$   &  $yes$  &  $yes$   \\
\hline\hline
\end{tabular}
\end{table}

\psubsubsection{Transverse wave}

For the force wave with angular frequency $\omega$, the force wave should be a harmonic function $cos \omega t$ ,  and then should be chosen as the function $exp(-i \omega t)$ . The force wave $\textbf{N}_e$ and $\textbf{N}_b$ can be written as follows,
\begin{eqnarray}
\textbf{N}_e = \textbf{N}_e (r) exp(- i \omega t ) ~,~ \textbf{N}_b = \textbf{N}_b (r) exp(- i \omega t ) ~,
\end{eqnarray}
combining this with Eq.(151), gives
\begin{eqnarray}
\left\{ - ( \omega / v_0 )^2 + \Sigma \partial_j^2 \right\} \textbf{N}_e (r) = 0~, ~ \left\{ - ( \omega / v_0 )^2 + \Sigma \partial_j^2 \right\} \textbf{N}_b (r) = 0~.
\end{eqnarray}

From the analysis of Eqs.(148) and (149), we find that the force wave should be one hyperbolic cosine $cos ( i \gamma)$ , and be chosen as the function $exp(-i \textbf{I} \gamma)$ . The wave-like vector $\textbf{K}_n = \Sigma (\emph{\textbf{I}}_j K_{nj})$ , while the $ \gamma = - \Sigma ( r_j K_{nj} ) $ , with $K_{nj}$ being the coefficient.

The force wave $\textbf{N}_e (r)$ and $\textbf{N}_b (r)$ are
\begin{eqnarray}
\textbf{N}_e (r) = \textbf{N}_{e0} \circ exp(-i \textbf{I} \gamma) ~,
~ \textbf{N}_b (r) = \textbf{N}_{b0} \circ exp(-i \textbf{I} \gamma) ~,
\nonumber
\end{eqnarray}
where $\textbf{N}_{e0}$ and $\textbf{N}_{b0}$ both are constant vectors in the octonion space.

And then we have the result from the above and Eq.(160) ,
\begin{eqnarray}
- (\omega / v_0 )^2 + \Sigma K_{nj}^2 = 0~.
\end{eqnarray}

In the electromagnetic field, we will find that force waves are the transverse waves in a vacuum. Similar to Eqs.(110)$\sim$(112), there are following results from Eqs.(148) and (149),
\begin{eqnarray}
& \textbf{K}_n \cdot \textbf{N}_{e0} = 0~, ~\textbf{K}_n \cdot \textbf{N}'_{e0} = 0~, ~\textbf{K}_n \cdot \textbf{N}_{b0} = 0~, ~\textbf{K}_n \cdot \textbf{N}'_{b0} = 0~,\\
& \textbf{K}_n \times \textbf{N}'_{e0} + k_v \omega \textbf{N}_{b0} = 0~,
~\textbf{K}_n \times \textbf{N}_{e0} - k_v \omega \textbf{N}'_{b0} = 0~,\\
&\textbf{K}_n \times \textbf{N}'_{b0} + \omega (\textbf{N}_{e0} / V_0 ) / v_0 = 0~,
~ \textbf{K}_n \times \textbf{N}_{b0} - \omega (\textbf{N}'_{e0} / V_0 ) / v_0 = 0~,
\end{eqnarray}
where $\textbf{N}'_{e0} = \textbf{N}_{e0} \circ \textbf{I}$ , $\textbf{N}'_{b0} = \textbf{N}_{b0} \circ \textbf{I}$ .

The above means that there are some relationships between $\mathbb{N}_e$ with $\mathbb{N}_e \circ \textbf{I}$ . The $\textbf{K}_n$ and $\textbf{N}_{e0}$ will yield a new component $\textbf{N}'_{b0}$ , while the $\textbf{K}_n$ and $\textbf{N}_{b0}$ produce the $\textbf{N}'_{e0}$ . All of these force components belong to the transverse waves.
Moreover, the amplitudes of force wave components, $\textbf{N}'_{e0}$ and $\textbf{N}'_{b0}$ , are the same as that of $\textbf{N}_{e0}$ and $\textbf{N}_{b0}$ respectively.

\psection{Flux waves in electromagnetic and gravitational fields}

From the property of octonion spaces, we can define the linear momentum flux $\mathbb{L}^p$, angular momentum flux $\mathbb{L}^l$, energy-torque flux $\mathbb{L}^w$, power-force flux $\mathbb{L}^n$, and physical quantity flux $\mathbb{L}^c$ respectively from the velocity $\mathbb{V}$, field potential $\mathbb{A}$, and linear momentum $\mathbb{P}$ etc.

Consequently there may exist some other kinds of wave equations in the electromagnetic and gravitational fields for those flux quantities.

\begin{table}[h]
\caption{Energy flux an others in the electromagnetic and gravitational fields.}
\label{tab:table1}
\centering
\begin{tabular}{lll}
\hline\hline
$linear~momentum~flux$                                & $angular~momentum~flux$                & $energy-torque~flux$     \\
\hline
$\mathbb{L}^p = (\mathbb{V} + k_{rx} \mathbb{A}) \circ \mathbb{P}$            & $\mathbb{L}^l = (\mathbb{V} + k_{rx} \mathbb{A}) \circ \mathbb{L}$   & $\mathbb{L}^w = (\mathbb{V} + k_{rx} \mathbb{A}) \circ \mathbb{W}$                           \\
$\mathbb{W}^p = v_0 ( \mathbb{B} /v_0 + \lozenge ) \circ \mathbb{L}^p$        & $\mathbb{W}^l = v_0 ( \mathbb{B} /v_0 + \lozenge ) \circ \mathbb{L}^l$             & $\mathbb{W}^w = v_0 ( \mathbb{B} /v_0 + \lozenge ) \circ \mathbb{L}^w$       \\
$\mathbb{N}^p = v_0 ( \mathbb{B} /v_0 + \lozenge )^* \circ \mathbb{W}^p$      & $\mathbb{N}^l = v_0 ( \mathbb{B} /v_0 + \lozenge )^* \circ \mathbb{W}^l$             & $\mathbb{N}^w = v_0 ( \mathbb{B} /v_0 + \lozenge )^* \circ \mathbb{W}^w$     \\
$\mathbb{C}^p = v_0 ( \mathbb{B} /v_0 + \lozenge )^* \circ \mathbb{N}^p$      & $\mathbb{C}^l = v_0 ( \mathbb{B} /v_0 + \lozenge )^* \circ \mathbb{N}^l$         & $\mathbb{C}^w = v_0 ( \mathbb{B} /v_0 + \lozenge )^* \circ \mathbb{N}^w$     \\
\hline\hline
\end{tabular}
\end{table}

To incorporate various linear momentum flux within a single definition, the linear momentum flux $\mathbb{L}^p$ is defined from the linear momentum density $\mathbb{P} = \mu_g \mathbb{S} / \mu_g$ in Eq.(24), the velocity $\mathbb{V}$, and the field potential $\mathbb{A}$ . The related equations can be written as follows,
\begin{eqnarray}
& \mathbb{L}^p = (\mathbb{V} + k_{rx} \mathbb{A} ) \circ \mathbb{P}~,~ \mathbb{W}^p = v_0 ( \mathbb{B}_g/v_0 + \lozenge) \circ \mathbb{L}^p~,\\
& \mathbb{N}^p = v_0 ( \mathbb{B}_g/v_0 + \lozenge)^* \circ \mathbb{W}^p~,~ \mathbb{C}^p = v_0 ( \mathbb{B}_g/v_0 + \lozenge)^* \circ \mathbb{N}^p~.
\end{eqnarray}

From the above, we can describe the property of linear momentum flux in the electromagnetic field in the presence of gravitational field.

We choose following definition of angular momentum flux to include various angular momentum flux in the electromagnetic field and gravitational field. In octonion space, the equations about the angular momentum flux are defined from the angular momentum density $\mathbb{L}$ , and the velocity $\mathbb{V}$ and field potential $\mathbb{A}$,
\begin{eqnarray}
& \mathbb{L}^l = (\mathbb{V} + k_{rx} \mathbb{A} ) \circ \mathbb{L}~,~ \mathbb{W}^l = v_0 ( \mathbb{B}_g/v_0 + \lozenge) \circ \mathbb{L}^l~,\\
& \mathbb{N}^l = v_0 ( \mathbb{B}_g/v_0 + \lozenge)^* \circ \mathbb{W}^l~,~ \mathbb{C}^l = v_0 ( \mathbb{B}_g/v_0 + \lozenge)^* \circ \mathbb{N}^l~.
\end{eqnarray}

Similarly the octonion energy-torque flux equations of electromagnetic field and gravitational field are defined as follows,
\begin{eqnarray}
& \mathbb{L}^w = (\mathbb{V} + k_{rx} \mathbb{A} ) \circ \mathbb{W}~,~ \mathbb{W}^w = v_0 ( \mathbb{B}_g/v_0 + \lozenge) \circ \mathbb{L}^w~,\\
& \mathbb{N}^w = v_0 ( \mathbb{B}_g/v_0 + \lozenge)^* \circ \mathbb{W}^w~,~ \mathbb{C}^w = v_0 ( \mathbb{B}_g/v_0 + \lozenge)^* \circ \mathbb{N}^w~.
\end{eqnarray}

The power-force flux and related equations in the octonion space can be defined from the power-force $\mathbb{N}$ ,
\begin{eqnarray}
& \mathbb{L}^n = (\mathbb{V} + k_{rx} \mathbb{A} ) \circ \mathbb{N}~,~ \mathbb{W}^n = v_0 ( \mathbb{B}_g/v_0 + \lozenge) \circ \mathbb{L}^n~,\\
& \mathbb{N}^n = v_0 ( \mathbb{B}_g/v_0 + \lozenge)^* \circ \mathbb{W}^n~,~ \mathbb{C}^n = v_0 ( \mathbb{B}_g/v_0 + \lozenge)^* \circ \mathbb{N}^n~.
\end{eqnarray}

To deduce the wave equation for $\mathbb{C}$ flux, we can similarly define following equations,
\begin{eqnarray}
& \mathbb{L}^c = (\mathbb{V} + k_{rx} \mathbb{A} ) \circ \mathbb{C}~,~ \mathbb{W}^c = v_0 ( \mathbb{B}_g/v_0 + \lozenge) \circ \mathbb{L}^c~,\\
& \mathbb{N}^c = v_0 ( \mathbb{B}_g/v_0 + \lozenge)^* \circ \mathbb{W}^c~,~ \mathbb{C}^c = v_0 ( \mathbb{B}_g/v_0 + \lozenge)^* \circ \mathbb{N}^c~.
\end{eqnarray}

From the above we will deduce some wave equations and transverse waves about the flux waves similar to that of Eqs.(67)$\sim$(164), including the inferences about the energy flux.

\begin{table}[h]
\caption{Some flux quantities in the electromagnetic and gravitational fields.}
\label{tab:table1}
\centering
\begin{tabular}{ll}
\hline\hline
$power-force~flux$       & $\mathbb{C}~flux$    \\
\hline
$\mathbb{L}^n = (\mathbb{V} + k_{rx} \mathbb{A}) \circ \mathbb{N}$        & $\mathbb{L}^c = (\mathbb{V} + k_{rx} \mathbb{A}) \circ \mathbb{C}$          \\
$\mathbb{W}^n = v_0 ( \mathbb{B} /v_0 + \lozenge ) \circ \mathbb{L}^n$    & $\mathbb{W}^c = v_0 ( \mathbb{B} /v_0 + \lozenge ) \circ \mathbb{L}^c$      \\
$\mathbb{N}^n = v_0 ( \mathbb{B} /v_0 + \lozenge )^* \circ \mathbb{W}^n$  & $\mathbb{N}^c = v_0 ( \mathbb{B} /v_0 + \lozenge )^* \circ \mathbb{W}^c$    \\
$\mathbb{C}^n = v_0 ( \mathbb{B} /v_0 + \lozenge )^* \circ \mathbb{N}^n$  & $\mathbb{C}^c = v_0 ( \mathbb{B} /v_0 + \lozenge )^* \circ \mathbb{N}^c$    \\
\hline\hline
\end{tabular}
\end{table}

Further, substituting $( \mathbb{D} + k_{rx} \mathbb{B})$ for $(\mathbb{V} + k_{rx} \mathbb{A})$ in Table 13 and Table 14, we will obtain some other sorts of equations similarly. Herein $( \mathbb{D} + k_{rx} \mathbb{B}) = \lozenge \circ(\mathbb{V} + k_{rx} \mathbb{A})$ .

\psection{CONCLUSIONS}

In the electromagnetic and gravitational fields, the field equations and their wave features can be rephrased with the algebra of quaternions and of octonions. The related conclusions include the wave equation, transverse wave, law of reflection, law of refraction, Fresnel formula, and total internal reflection, etc.

With the algebra of quaternions, we can draw out the wave equation from the field equation in the gravitational field, and realize that the gravitational waves belong to transverse waves. Introducing the quaternion exponential function $ exp( i \textbf{i} \alpha )$ , the wave vector and gravitational wave, $\textbf{g}_0$ and $\textbf{b}_0$ , will produce two new wave components, $\textbf{g}'_0$ and $\textbf{b}'_0$ , respectively. Moreover, the $\textbf{g}_0$ and $\textbf{b}'_0$ can be determined simultaneously, while the $\textbf{g}'_0$ and $\textbf{b}_0$ can be at the same time.

In the electromagnetic field, making use of the algebra of octonions, we can deduce the wave equation, laws of reflection and refraction, Fresnel formula, and total internal reflection, and find that electromagnetic waves belong to the transverse wave in a vacuum. Bringing in the octonion exponential function $exp( i \textbf{I} \alpha )$, the wave vector and electromagnetic wave components, $\textbf{E}_0$ and $\textbf{B}_0$, will produce two new wave components, $\textbf{E}'_0$ and $\textbf{B}'_0$ , respectively. In contrast to the conventional electromagnetic theory with vector terminology, the research points out that the $\textbf{E}_0$ and $\textbf{B}_0$ can not be determined simultaneously via the wave vector. But the $\textbf{E}_0$ and $\textbf{B}'_0$ can be measured at the same time, while the $\textbf{E}'_0$ and $\textbf{B}_0$ can also be.

Similar to the field strength, the force and torque etc may possess respectively the static quantity, alternating quantity, and independent wave etc. Besides the wave equation for field strength, there are also the wave equations for force and torque in the gravitational field, and that for force-like and torque-like in the electromagnetic field. In the octonion space, the mass continuity equation as well as the charge continuity equation are the gauge equations for Maxwell-like equations of torque wave and of torque-like wave respectively. The angular frequency $\omega$ of gravitational wave may be different to which of electromagnetic wave, force wave, force-like wave, torque wave, or torque-like wave etc. Similarly the speed $v_0$ of $\textbf{g} / v_0$ may be diverse from that of $\textbf{w}_e / v_0$ or $\textbf{n}_e / v_0$ etc, while the speed $V_0$ of $\textbf{E} / V_0$ may be distinct from that of $\textbf{W}_e / V_0$ or $\textbf{N}_e / V_0$ etc.

It should be noted that the study for the deductions of wave equations in the electromagnetic fields and gravitational fields examined only some simple cases in quaternion spaces and octonion spaces. Despite its preliminary characteristics, this study can clearly indicate that the wave equation of electromagnetic field can be deduced from the field equation with the algebra of octonions, and obtain some inferences such as Fresnel formula, law of reflection, law of refraction, and total internal reflection etc in the vacuum far away from the field sources. For the future studies, the research will concentrate on only the electromagnetic waves transmitting in the conductors.

\ack
This project was supported partially by the National Natural Science Foundation of China under grant number 60677039.

\end{paper}

\end{document}